\documentclass[preprint,nofootinbib,aps,showpacs,preprintnumbers]{revtex4-1}

\usepackage{graphicx}
\usepackage{dcolumn}
\usepackage{appendix}
\usepackage{epsfig}
\usepackage{epstopdf}
\usepackage{amssymb}
\usepackage{amsfonts}
\usepackage{xcolor}
\usepackage{float}
\usepackage{hyperref}
\usepackage{physics}
\usepackage{comment}
\usepackage[caption=false]{subfig}
\usepackage{soul}
\usepackage[bottom]{footmisc}

\usepackage{lineno}
\usepackage{amsmath}
\usepackage{dirtytalk}

\hypersetup{
	colorlinks = true,
	linkcolor = blue,
	citecolor = blue,
	urlcolor = blue
}
\allowdisplaybreaks
\setstcolor{red}

\begin{document}

\title{Non-resonant inter-species interaction and its effect on the position response function of cold atoms}

\author{Anirban Misra$^1$}
\email{anirbanm@rrimail.rri.res.in}
\author{Urbashi Satpathi$^2$}
\email{urbashi.satpathi@gmail.com}
\author{Supurna Sinha$^1$}
\email{supurna@rri.res.in}
\author{Sanjukta Roy$^1$}
\email{sanjukta@rri.res.in}
\author{Saptarishi Chaudhuri$^1$}
\email{srishic@rri.res.in}

\affiliation{$^1$ Raman Research Institute, C. V. Raman Avenue, Sadashivanagar, Bangalore-560080, India.}
\affiliation{$^2$Department of Physics and Materials Science Engineering, Jaypee Institute of Information Technology, A-10, Sector 62, Noida, UP-201309, India.}

\date{\today}
\begin{abstract}
\noindent In the context of non-equilibrium statistical physics, the position response of a particle, coupled to a bath, subjected to an external force is a topic of broad interest. A topic of further interest is two distinguishable sets of interacting particles in contact with two different baths. Here, we report the experimental evidence of the modification of the position response function (PRF) of an ensemble of cold atoms in a magneto-optical trap when it is placed alongside a dilute cloud of cold atoms of a different species.  Our experiment consists of a mass-imbalanced cold atomic mixture of Potassium and Sodium atoms. We focus on the position response of Potassium atoms when subjected to a sudden displacement in the presence of a cold Sodium atomic cloud. Notably, we find that, in the underdamped regime of motion, the oscillation frequency of motion of the cold atoms changes as much as 30 $\%$ depending on the effective inter-species light-assisted interaction strength. On the other hand, in the overdamped regime, there is a reduction, as high as 10.5 $\%$ in the damping coefficient, depending on the interaction strength. Using a quantum Langevin approach, we develop a framework that aligns well with experimental results, with potential applications in mass and charge transport studies under varied physical conditions simulated in cold atoms.
\end{abstract}

\maketitle

\section{\label{sec:level1}Introduction}
Cold atoms provide an ideal test bed for precision measurements of the response to external perturbations \cite{morawetz2014universal, inguscio2013atomic, bhar2022measurements, misra2024effect}. With many experimental systems simultaneously trapping and cooling neutral atoms, inter-species interaction-dependent trap loss spectroscopy has been a subject of intense study in the recent past \cite{sutradhar2023fast,hewitt2024controlling,bhatt2022stochastic}. Similar studies have been performed in ion-atom systems \cite{dieterle2020inelastic,xing2022ion, weckesser2021observation,RMPTomza2019}, atom-molecule systems \cite{Gregory_2021,Pan2024,defenu2023long} and Rydberg atom - neutral atom systems \cite{Pfau2023atom-ion,wang2020optical,hirzler2021rydberg,cheng2024emergent}. These studies aim at the understanding of collision and scattering properties at ultra-low temperatures as well as an exploration of exotic quantum phases of matter in the presence of interactions. However, there is also a fundamental effect on modification of the position response of atoms even when the temperature of the cloud is not below the critical temperature for quantum phase transition. We have shown in a recent study, how the presence of intra-species interactions ($\sim\frac{1}{R^3}$ where $R$ is the distance between the atoms) results in a significant change in the nature of the cold atom position response function \cite{misra2024effect}.  In this paper, we focus on the effect of inter-atomic interaction and discuss the response function of cold $^{39}$K atoms in the presence of $^{23}$Na atoms in a magneto-optical trap (MOT) in a temperature regime ranging from 250 $\mu K$ - 1 $mK$.  This experiment shows how the position response function (PRF) of a particular species in the presence of another species can serve as a probe for a relatively weaker interaction in typical interatomic separation in a MOT ($\sim\frac{1}{R^6}$ where $R$ is the distance between the atoms of different species, typically ranging from 5 $\mu m$ - 10 $\mu m$ on an average). By suitably selecting the trapping laser and magnetic field parameters, we can conduct our experiments both in the underdamped and overdamped regimes of motion \cite{bhar2022measurements,misra2024effect}. We explore the modification of motion of atoms in both regimes in the presence and absence of another species of cold atoms. In this case the atoms of different species are in separate photon baths created by the optical molasses. The inter- and intra-species interactions between the cold atoms happen through light-assisted collisions. Furthermore, our theoretical model based on the generalised quantum Langevin equation gives a qualitative as well as a quantitative explanation of the results of the experiment.  We use a system-bath model\cite{Caldeira1983,PhysRevA.37.4419,ford1988independent,bhar2022measurements,misra2024effect} for the two species cold atomic clouds and obtain an effective description of the dynamics of the centroids of the clouds which leads to an expression
for the position response function. We use a combination of analytical and numerical calculations to theoretically study the variation of the response function with the strength of the inter-species interaction in the overdamped and underdamped regimes and compare against experimental results. We find good qualitative and quantitative agreement between theory and experiment. 

\section{\label{sec:level2}Interaction dependence of the cold-atom position response function}
\subsection{\label{subsec:level2}The Experimental system}
\begin{figure}[htbp]
    \centering
    \includegraphics[width=0.7\linewidth]{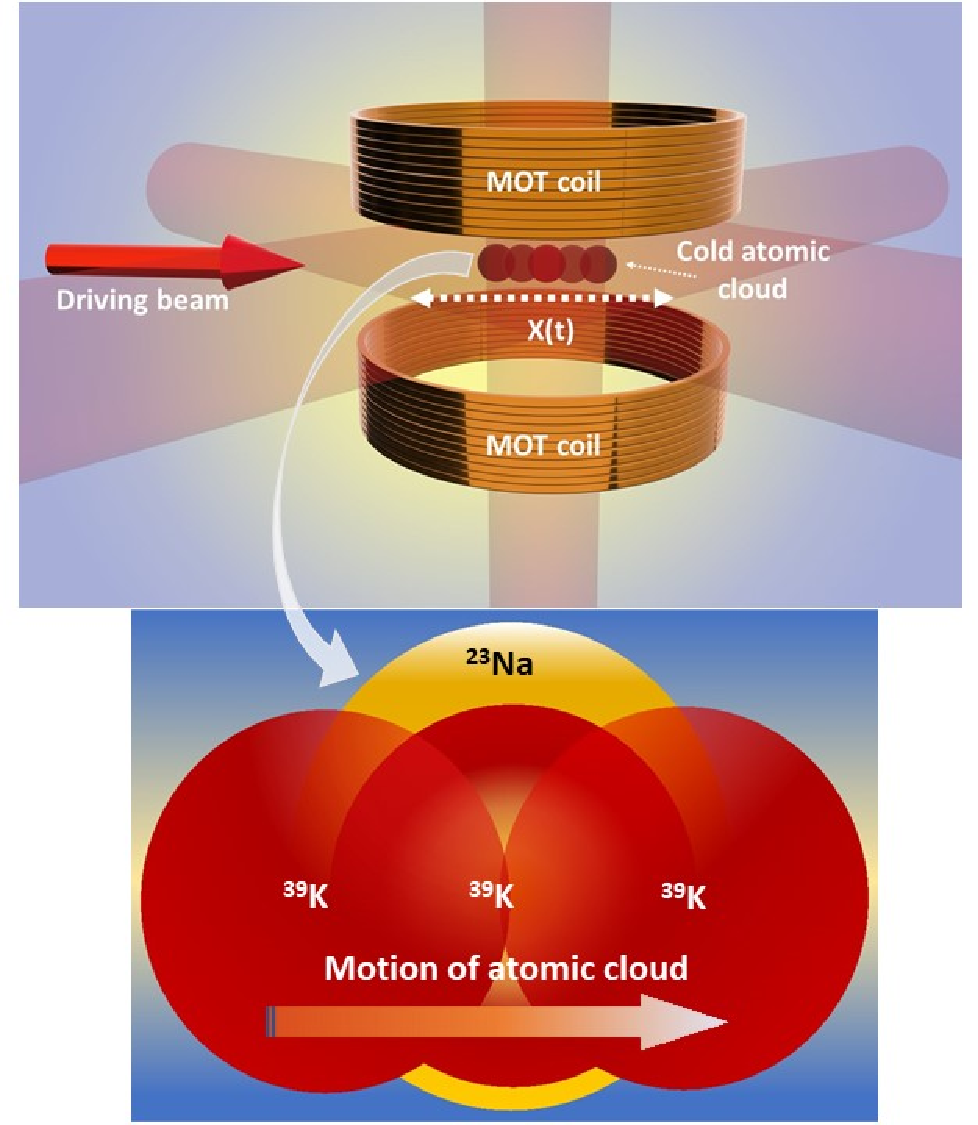}
    \caption{A schematic diagram of the experimental set-up. The driving beam is directed towards the trapped cold $^{39}$K atomic cloud and it perturbs the atoms from their initial positions. The position response function (PRF) of the perturbed cold $^{39}$K atoms is recorded both in the absence and in the presence of the trapped cold $^{23}$Na atoms.}
    \label{fig:enter-label}
\end{figure}
\hspace{-0.35cm}A detailed description of our apparatus, designed to simultaneously capture substantial number of $^{39}$K and $^{23}$Na atoms within a Magneto-Optical Trap (MOT), can be found in \cite{sutradhar2023fast}. Our system facilitates a rapid (within a second) loading of atoms into a 3D Magneto-Optical Trap (3DMOT) through the use of two spatially separated 2D Magneto-Optical Traps (2D$^{+}$MOTs). This configuration allows us to control the number of atoms trapped in the MOT of each species individually in a fast and controlled manner. Also, our experiment is able to record fluorescence images with significantly enhanced signal-to-noise ratio (Typically $\sim1600$), which allows for accurate determination of the atomic cloud's centroid position ($\sim\pm 5$ $\mu$m). Moreover, the background vacuum is maintained at low $10^{-11}$ mbar level which allows the trap lifetime to be approximately 40 seconds. 
\par
In the set of experiments described in the latter part of this paper, the cloud size ($1/e^{2}$ radius) for $^{39}$K atoms was varied from 0.7 $mm$ to 2.1 $mm$, while keeping the sodium cloud size constant at 0.7 $mm$. The number of atoms in the $^{39}$K cold cloud has been varied from $\sim5\times10^7$ to $\sim1\times10^9$ while the $^{23}$Na cloud atom number has been kept constant at $\sim1\times10^7$.
\par
After capturing the dual species cold atom clouds in the 3DMOT with the desired size and atom number ratios and reaching an equilibrium condition, we pulse an additional weak laser beam, the `driving-beam', onto the trapped $^{39}$K atomic cloud to induce a sudden displacement of the cold $^{39}$K atoms from the trap's center. The intensity and the detuning from the $^{39}$K $D2$ transition frequency of this driving beam is carefully chosen so that it does not induce significant atom-loss. After the $^{39}$K atomic cloud reaches its equilibrium position under the influence of the driving beam, its centroid position is tracked using time-delayed fluorescence images. \textbf{Fig-\ref{fig:enter-label}} is a schematic diagram of the experiment and the lower cartoon indicates the movement of the  $^{39}$K atomic cloud in the presence of the $^{23}$Na atomic cloud. Notice that we can selectively drive the $^{39}$K atoms without affecting the $^{23}$Na atomic cloud even slightly. This allows us to study the inter-species interaction dominated response function in a controlled manner.
\par
\begin{figure}[htbp]
    \centering
    \includegraphics[width=\linewidth]{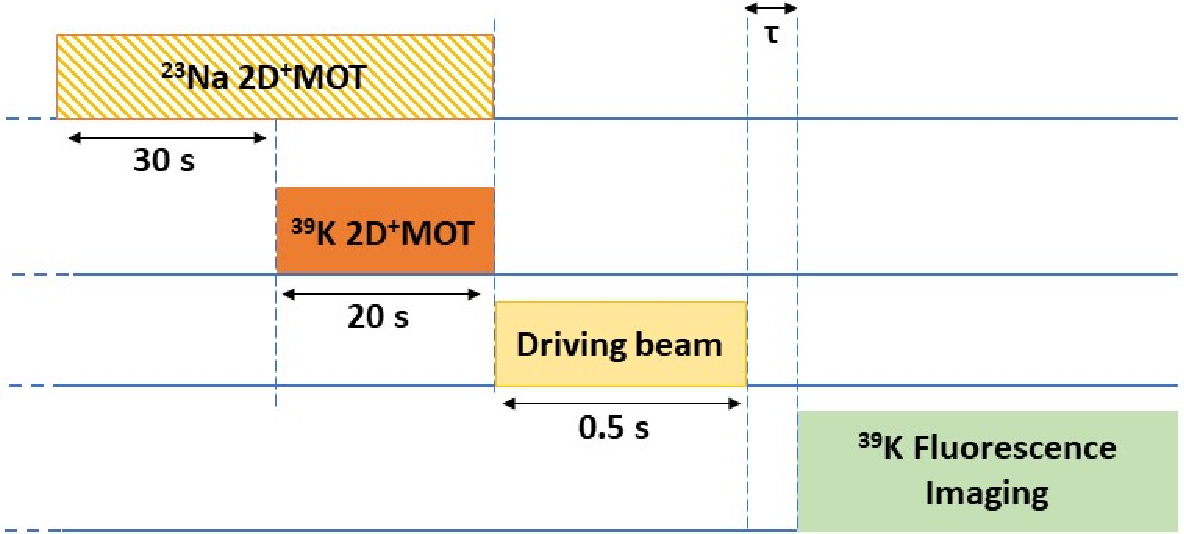}
    \caption{Time sequence of the experiment conducted in two parts. In the first part, $^{39}$K atoms are loaded and spatially perturbed using a driving beam. The resulting cloud position is analyzed to determine the response function of $^{39}$K atoms. In the second part, both $^{23}$Na and $^{39}$K atoms are sequentially loaded into the MOT (by keeping their respective 2D$^{+}$MOTs ON). The position response of $^{39}$K atoms is then measured in the presence of $^{23}$Na atoms. The post-drive delay ($\tau$) is identical for both parts to enable direct comparison.}
    \label{fig:Time-sequence}
\end{figure}
Notably, our experimental approach is divided into two distinct temporal phases as depicted in \textbf{Fig-\ref{fig:Time-sequence}}. During the initial phase of the experiment, only $^{39}$K atoms are loaded into the 3DMOT, and the positional response function of these atoms is analyzed after subjecting them to driving. In the subsequent phase, both $^{39}$K and $^{23}$Na atoms are introduced into the MOT, and we again investigate the positional response function specifically for the $^{39}$K atoms. Through analysis of the collected fluorescence images which shows the atomic density profile of the clouds, we are able to discern the differences in the response functions obtained from these two distinct experiments performed under otherwise identical conditions, thereby providing insights into the interaction and behavior of the atoms within the trap. The fluorescence from the $^{39}$K atoms are collected using an ICCD camera (Andor SOLIS ICCD-3569) with an interference filter allowing only fluorescence from the $^{39}$K cloud. In a typical experimental sequence, the fluorescence image is taken after displacement and a variable time-delay with an exposure time of 1 $ms$. Long exposure time allows us to collect enough fluorescence photons for accurate determination of the density profile. On the other hand, the exposure time needs to be as short as possible, so that the instantaneous position of the cloud can be determined for further analysis. We found an exposure time of 1 $ms$  to be optimized for satisfying these two contradictory requirements. After the fluorescence image is collected, the 3DMOT is switched OFF by switching off the cooling laser beams and the trapping magnetic field. The entire sequence is repeated multiple times for statistically significant datasets.
\par
In our previous study \cite{sutradhar2023fast}, we found that introducing $^{39}$K atoms into a magneto-optical trap (MOT) already containing cooled and trapped $^{23}$Na atoms led to a decrease in the number of trapped $^{23}$Na atoms. To address this issue in the second part of our experimental procedure, we first load the $^{23}$Na atoms into the MOT. We continue to do so until the fluorescence signal from the $^{23}$Na atoms on the camera reaches saturation, which takes approximately 30 seconds. At this point, we begin to load the $^{39}$K atoms into the MOT. We then allow about 20 seconds for the $^{23}$Na atom count to stabilize before we start capturing images to analyze the dynamics of the $^{39}$K atoms, due to the effect of the driving beam.

\subsection{\label{subsec:level2.1}Theoretical framework}
\begin{figure}[htbp]
    \centering
    \includegraphics[width=0.8\linewidth]{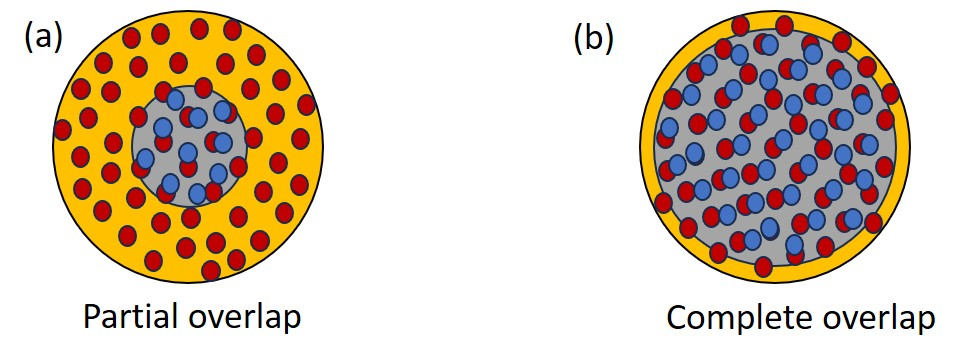}
    \caption{A schematic diagram of the overlapping clouds of the two species. The fixed yellow cloud with red particles is the $^{23}$Na cloud and the varying gray cloud with blue particles is the $^{39}$K cloud. (a) The case where the clouds are partially overlapping. (b) The case where the clouds are completely overlapping. In (a), both $N_1,N_2$ are very small in the interacting or overlapping region making the interaction term $N_1N_2l_2$ very small, in (b), both $N_1,N_2$ are very large in the overlapping region making the interaction term $N_1N_2l_2$ very large.}
    \label{fig:Modelfig}
\end{figure}

We consider clouds of two species interacting with one another and each interacting with a separate thermal bath. Schematic diagrams of the clouds of the two species overlapping with one another (partially and completely) are shown in \textbf{Fig-\ref{fig:Modelfig}}. 
Let us consider that species one has $N_1$ particles with mass $M_1$, position and momentum coordinates $X_j,P_j$, $j=1,2..,N_1$ and the second species has $N_2$ particles with mass $M_2$, position and momentum coordinates $Y_k, Q_k$, $k=1,2..,N_2$, in the overlapping region of the two clouds. $\omega_1$ and $\omega_2$ are the trapping frequency of the first species and the second species respectively. The total Hamiltonian of the two species system interacting with two distinct baths, is given by,
\begin{align}
    H&=\sum_{j=1}^{N_1}\left(\frac{P_j^2}{2M_1}+\frac{1}{2}M_1 \omega_1^2 X_j^2 \right)+\sum_{k=1}^{N_{2}}\left(\frac{Q_k^2}{2M_2}+\frac{1}{2}M_2 \omega_2^2 Y_k^2\right)\nonumber\\&-l_1\left(\sum_{j,s,j\neq s}^{N_1} X_j X_s+\sum_{k,l,k\neq l}^{N_2} Y_k Y_l\right)-l_2\sum_{j=1}^{N_1} X_j\sum_{k=1}^{N_2} Y_k +\nonumber\\&\sum_i \left(\frac{p_{1i}^2}{2 m_{1i}}+\frac{1}{2}m_{1i}\omega_{1i}^2 x_{1i}^2\right)+\sum_i \left(\frac{p_{2i}^2}{2 m_{2i}}+\frac{1}{2}m_{2i}\omega_{2i}^2 x_{2i}^2\right)\nonumber\\&-\sum_i \left(\lambda_{1i}x_{1i}\sum_j X_j + \frac{\lambda_{1i}^2}{2 m_{1i}\omega_{1i}^2}\sum_j X_j^2\right)\nonumber\\&-\sum_i \left(\lambda_{2i}x_{2i}\sum_k Y_k + \frac{\lambda_{2i}^2}{2 m_{2i}\omega_{2i}^2}\sum_k Y_k^2\right)
\end{align}
Here, $l_1=\frac{2q^2}{4\pi\epsilon_0 R^3}$ and $l_2=\frac{2 q^2 I_1 I_2}{E_1 E_2 (I_1+I_2)(4\pi\epsilon_0)^2R^6}$ are the dipole-dipole intra-species interaction strength and the dipole-induced dipole inter-species interaction strength respectively \cite{jackson2012classical}. $R$ is the inter-atomic separation, $I_j$  and $E_j$ $j=1,2$ are the Ionisation energies and electric fields of the two species. 
$p_{1i}$, $x_{1i}$, ${ m_{1i}}$, $\omega_{1i}$, $\lambda_{1i}$ and $p_{2i}$, $x_{2i}$, ${ m_{2i}}$,  $\omega_{2i}$, $\lambda_{2i}$
are respectively the momentum coordinate, the position coordinate,  the mass, the angular frequency and  the coupling constant 
pertaining to the first bath and the second bath. The bath particles do not interact with one another but only with the system centroid coordinates. $X_c=\sum_j X_j, Y_c=\sum_k Y_k$ are the centroid of the first and second species respectively. 
In order to get the Quantum Langevin Equation (QLE) for each species, we write the Heisenberg equations of motion for the system and the bath coordinates \cite{misra2024effect}. After integrating out the bath degrees of freedom we get the QLEs for the centroids of the two species,
\begin{align}
    M_1 \ddot{X_c}(t)&=-M_1 \omega_1^2 X_c(t) +l_1 X_c(t) (N_1-1)+l_2 Y_c(t) N_1 \nonumber\\&+N_1 \zeta_1(t) - N_1\gamma_1(t) \dot{X_c}(t)\\
   M_2 \ddot{Y_c}(t)&=-M_2 \omega_2^2 Y_c(t) +l_1 Y_c(t) (N_2-1)+l_2 X_c(t) N_2 \nonumber\\&+ N_2 \zeta_2(t) - N_2\gamma_2(t) \dot{Y_c}(t) 
\end{align}

$\zeta_i, \gamma_i, i=1,2$ are respectively the noise and dissipation terms for the two thermal baths and are given by,
\begin{align} 
\gamma_1(t)&=\sum_i \frac{\lambda_{1i}^2}{m_{1i}\omega_{1i}^2}\cos{\omega_{1i}(t)}\\
 \zeta_1(t)&=\sum_i \lambda_{1i}\left[\left(x_{1i}(0)-\frac{\lambda_{1i}}{m_{1i}\omega_{1i}^2}X_c(0)\right)\cos{\omega_{1i}}(t)\right.\nonumber\\&+\left.\frac{p_{1i}(0)}{m_{1i}\omega_{1i}}\sin{\omega_{1i}}(t)\right]\\
 \gamma_2(t)&=\sum_i \frac{\lambda_{2i}^2}{m_{2i}\omega_{2i}^2}\cos{\omega_{2i}(t)}\\
 \zeta_2(t)&=\sum_i \lambda_{2i}\left[\left(x_{2i}(0)-\frac{\lambda_{2i}}{m_{2i}\omega_{2i}^2}Y_c(0)\right)\cos{\omega_{2i}}(t)\right.\nonumber\\&+\left.\frac{p_{2i}(0)}{m_{2i}\omega_{2i}}\sin{\omega_{2i}}(t)\right]
\end{align}
 Defining, $M_1\omega_1^2 -l_1 (N_1-1)=W_1^2, M_2\omega_2^2 -l_1 (N_2-1)=W_2^2$, we can express the above QLEs as follows : 
\begin{align}
M_1 \ddot{X_c}(t)&=-W_1^2 X_c(t) +l_2 Y_c(t) N_1 + N_1 \zeta_1(t) - N_1\gamma_1(t) \dot{X_c}(t)\\
M_2 \ddot{Y_c}(t)&=-W_2^2 Y_c(t) +l_2 X_c(t) N_2 + N_2 \zeta_2(t) - N_2\gamma_2(t) \dot{Y_c}(t) 
\end{align}
After Fourier transforming, we get the following solutions to these QLEs,
\begin{align}
    X_c(\omega)&=\frac{N_1 D_2 \zeta_1(\omega)}{D}+\frac{N_1 N_2 l_2 \zeta_2(\omega)}{D}\\
    Y_c(\omega)&=\frac{N_2 D_1 \zeta_2(\omega)}{D}+\frac{N_1 N_2 l_2 \zeta_1(\omega)}{D}
\end{align}
Here,
$$D_j=-M_j^2 \omega^2 +W_j^2-i\omega N_j\gamma_j, j=1,2, \text{and} \;D=D_1 D_2-N_1 N_2 l_2^2$$
Using the Fluctuation Dissipation Theorem connecting the correlation and position response function, and, noise correlation and the dissipation terms in the Fourier domain, we get the expressions for the position response functions (PRF) for the two species,
\begin{align}
    \Im R_{X_c}(\omega)&=\frac{\omega\left(N_1^2 D_2 D_2^* \Re \gamma_1(\omega)+N_1^2 N_2^2 l_2^2 \Re\gamma_2(\omega)\right)}{D D^*}\label{ImRx}\\
   \Im R_{Y_c}(\omega)&=\frac{\omega\left(N_2^2 D_1 D_1^* \Re \gamma_2(\omega)+N_1^2 N_2^2 l_2^2 \Re\gamma_1(\omega)\right)}{D D^*} \label{ImRy}
\end{align}
As in \cite{misra2024effect} we consider an Ohmic bath model for the baths, i.e. $\gamma_1(\omega)=\gamma_1$ and $\gamma_2(\omega)=\gamma_2$. For the Ohmic baths we numerically calculate the imaginary parts of the PRFs. Using Kramers-Kronig relation we calculate the real parts of the PRFs.
After obtaining the imaginary and real parts, we calculate the PRFs of both species.
\begin{align}
  R_{X_c}(t)&=\frac{1}{2\pi}\int_{-\infty}^\infty \left(\Re R_{X_c}(\omega)+i\Im R_{X_c}(\omega)\right) e^{-i\omega t}d\omega\label{RXct}\\
  R_{Y_c}(t)&=\frac{1}{2\pi}\int_{-\infty}^\infty \left(\Re R_{Y_c}(\omega)+i\Im R_{Y_c}(\omega)\right) e^{-i\omega t}d\omega\label{RYct}
\end{align}
The theoretical analysis enables us to study the PRFs for both species for various bath models. However, we restrict ourselves to the Ohmic bath model to compare our theoretical results with the experimental observations.
This choice is motivated by the optical molasses in the cold-atom experimental setup. In fact, an Ohmic bath is equivalent to a force proportional to the velocity with a fixed coefficient of proportionality or damping coefficient, consistent with the force-velocity relation seen for an atom in optical molasses. 
To compare with the experimental results, we consider $^{39}$K to be the first species (the gray cloud with varying size) and $^{23}$Na to be the second species (the fixed yellow cloud) (See \textbf{Fig-\ref{fig:Modelfig}}). 
The PRFs for $^{39}$K and $^{23}$Na are given by Eqs. (\ref{RXct}) and (\ref{RYct}). For $l_2=0$, i.e. in the absence of inter-species interaction, the PRFs corresponds to the one pertaining to a single species:
\begin{align}
    R_{X_c}(t)|_{l_2 =0}=\frac{N_1}{M_1 \gamma_{1c}}e^{-\gamma_{11} t}\sinh{\gamma_{1c}t}\\
    R_{Y_c}(t)|_{l_2 =0}=\frac{N_2}{M_2 \gamma_{2c}}e^{-\gamma_{22} t}\sinh{\gamma_{2c}t}
\end{align}
where, $\gamma_{11}=\frac{N_1\gamma_1}{2 M_1}$, $\gamma_{22}=\frac{N_2\gamma_1}{2 M_2}$, $\gamma_{1c}=\sqrt{\gamma_{11}^2-W_{11}^2}$, $\gamma_{2c}=\sqrt{\gamma_{22}^2-W_{22}^2}$ and $W_{11}=W_1/\sqrt{M_1}$, $W_{22}=W_2/\sqrt{M_2}$. $\gamma_{11},\gamma_{22}$ are the damping coefficients and $W_{11}, W_{22}$ are the oscillation frequencies of $^{39}$K (first species) and $^{23}$Na (second species), respectively. The sign of $\gamma_{1c}$ and $\gamma_{2c}$ controls the nature of the PRFs. Note that in the experiment, we analyze the dynamics of the first species $^{39}$K only . So for comparison with the experimental data we will discuss only the PRF of $^{39}$K i.e. $R_{X_c}(t)$ in this section.
If $\gamma_{1c}$ is real then the system is in the overdamped regime where, $\gamma_{11}>W_{11}$, the damping coefficient dominates over the oscillation frequency and we notice an overdamped  behaviour of the PRF $R_{X_c}(t)|_{l_2 =0}$. If $\gamma_{1c}$ is imaginary, then the system is in the underdamped regime where $\gamma_{11}<W_{11}$ and the oscillatory frequency dominates over the damping coefficient and the dynamics of the PRF exhibits
a damped oscillatory behavior.
For finite $l_2$, the deviation from the single species PRF behaviour depends on the second term of Eq. \ref{ImRx} which is a function of $N_1N_2l_2$. In order to understand this deviation we can do a Binomial expansion considering ${\frac{N_1N_2l_2^2}{D_1D_2}}$ and ${\frac{N_1N_2l_2^2}{D_1^*D_2^*}}$ as small parameters, and truncating the expansion to the first order terms only. Hence the time dependent PRF of $^{39}$K reduces to,


\begin{align}
   R_{X_c}(t) = & R_{X_c}(t) \big|_{l_2 = 0} + \frac{N_1^2 N_2 l_2^2}{4M_1^2 M_2} 
   \Re\Bigg[\frac{e^{-\gamma_{11}t}}{Dc_1}\Big(\gamma_{11}\gamma_{1c} C_1 \cosh(\gamma_{1c}t) + C_2 \sinh(\gamma_{1c}t)\Big) \nonumber \\
   & + \frac{4 e^{-\gamma_{22}t}}{Dc_2}\Big(\gamma_{2c} C_3 \cosh(\gamma_{2c}t) - \gamma_{22} C_4 \sinh(\gamma_{2c}t)\Big)\Bigg] 
   \label{Rtfinal}
\end{align}

The details of the calculations which lead to Eq. (\ref{Rtfinal}) and the functional forms of $C_1, C_2, C_3, C_4$, $Dc_1, Dc_2$ are discussed in the Appendix \cite{Appendix}.

When the overlap of the clouds is very small, both $N_1$ and $N_2$ are very small numbers. Hence the correction terms obtained by the Binomial expansion  are negligibly small as $N_1N_2l_2$ is exceedingly small and hence the deviation of the two species PRF from the leading term which is the single species PRF of $^{39}$K, is hardly perceptible. As the overlap increases, both $N_1$ and $N_2$ increase, and thus the correction terms also have considerable contribution and we see a significant difference in the behaviour of the PRF compared to the leading term which pertains to the single species case both in the overdamped and underdamped regimes. Notice that the correction terms also involve the hyperboic cosine and sine terms which exhibit oscillatory or damped behaviour depending on the relative strengths of $\gamma_{11}$ and $W_{11}$ and $\gamma_{22}$ and $W_{22}$. So, we can clearly see that inter-species interaction controls the qualitative nature of the PRF. In the next section we discuss the PRF for both the overdamped and underdamped regimes. We plot the normalised PRF, which is $R_{X_c}(t)$ divided by the maximum value of $R_{X_c}(t)$ in the figures displayed (\textbf{Fig-4} and \textbf{Fig-6}). We have dropped the subscript $X_c$ and used $R(t)$ as a notation for PRF of $^{39}$K for the plots and discussions, since we will be discussing only the first species, i.e $^{39}$K species dynamics in the presence of $^{23}$Na cloud. 
\section{Experimental results and comparison with theory}
\begin{figure}[htbp] 
    \centering
    \includegraphics[width=\textwidth]{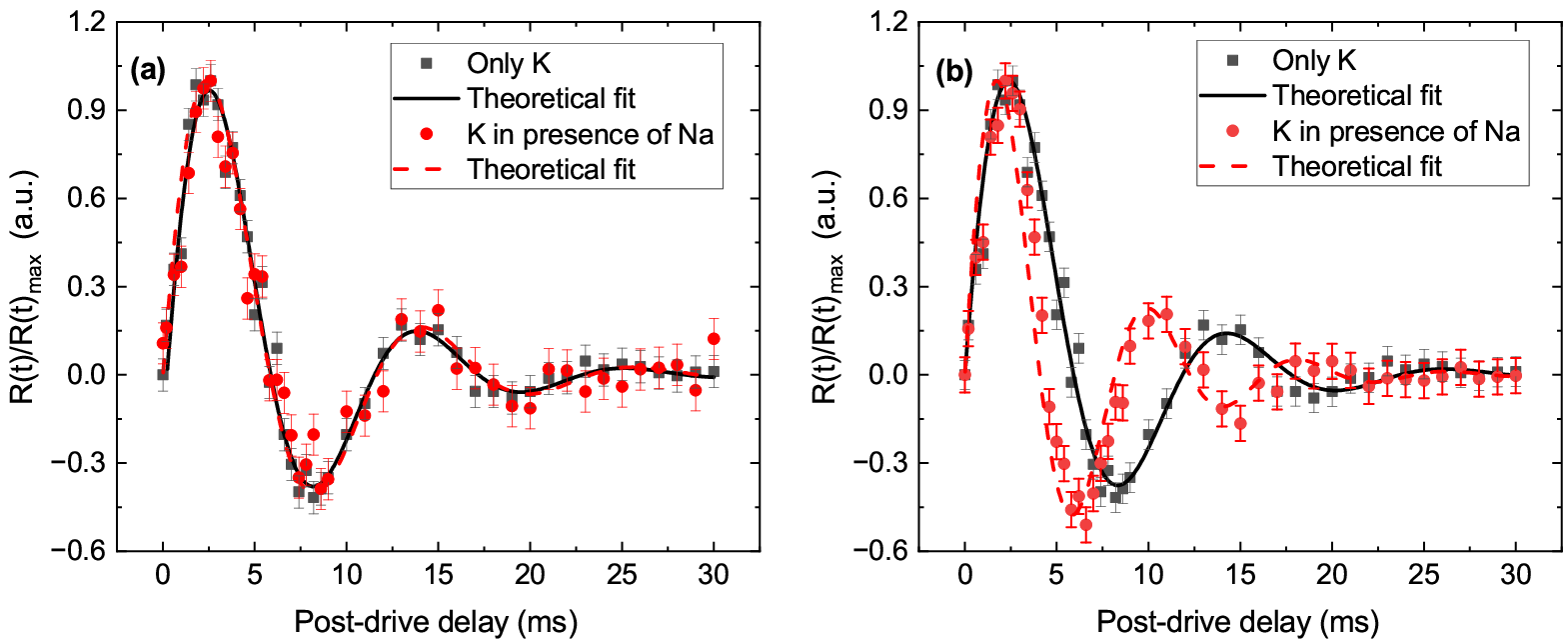}
    \caption{The response function of the $^{39}$K atoms due to the perturbation with a driving beam. The two different sets of data points in the graphs (circular and square) depict the response function of the $^{39}$K atoms in the presence and in the absence of the $^{23}$Na atoms respectively in the same Magneto-optical trap. The data in the graph has been taken when the cold atomic clouds of $^{39}$K and $^{23}$Na have a partial overlap in their unperturbed positions. As one can observe in sub-figure \textbf{(a)} the frequency of the PRF is the same in both scenarios for the $^{39}$K atom when the overlap is partial. On the other hand, the frequency of oscillation increases when the two clouds are almost equal in size and the clouds have a complete overlap. In both cases we have compared against the graphs (black solid lines and red dashed lines) stemming from our theoretical analysis. The parameters for the theoretical fit of normalized PRF of only $^{39}$K cloud are, $N_1 N_2 l_2=0$, $\gamma_{11}=165$ Hz, $ \gamma_{1c}=556 i$ Hz for both (a) and (b); and of $^{39}$K cloud in the presence of $^{23}$Na cloud in case (a) $N_1 N_2 l_2=200$, $\gamma_{11}=161$ Hz, $\gamma_{22}=140$ Hz, $\gamma_{1c}=545 i$Hz and $\gamma_{2c}=500 i$ Hz and in case (b) $N_1 N_2 l_2=980000$, $\gamma_{11}=181$ Hz, $\gamma_{22}=160$ Hz, $\gamma_{1c}=769.2 i$ Hz and $\gamma_{2c}=724.2 i$ Hz.}
    \label{fig:Fully_overlapped_UD}
\end{figure}

First, we experimentally investigate the position response function of $^{39}$K atoms in the underdamped regime. We obtain an oscillatory position response of the $^{39}$K cold atomic cloud when the magnetic field gradient is $\sim$ 18 $G/cm$ and the cooling beam intensity is $\sim$ 13.3 $I_{sat}$ (per beam). By increasing the MOT magnetic field gradient and/or decreasing the molasses intensity we can control the frequency and damping coefficient of the oscillatory cloud. However, the optimized parameters are chosen also to keep the atom number in the cold cloud high while maintaining the underdamped motion. A typical position response as a function of time is shown in \textbf{Fig-\ref{fig:Fully_overlapped_UD}(a)} (the black squares) and also in \textbf{Fig-\ref{fig:Fully_overlapped_UD}(b)} (the black squares). From this data we obtain a typical oscillation frequency of the cold atoms around 90.6 $\pm$ 0.5 $Hz$.

We observe a considerable change in the dynamics of the $^{39}$K atomic cloud with the change in the extent of overlap and the relative size of the $^{23}$Na and $^{39}$K atomic clouds. We systematically study the effect of the spatial overlap between the clouds of both species on the position response function of the $^{39}$K atoms. When the $^{23}$Na and $^{39}$K clouds are of comparable size and there is a considerable overlap of the clouds, as shown in\textbf{ Fig-\ref{fig:Fully_overlapped_UD}(b)}, we observe an increase in the frequency of the position response function (PRF) of the $^{39}$K atoms. In contrast, in \textbf{Fig-\ref{fig:Fully_overlapped_UD}(a)}, where the size of the $^{39}$K  cloud is much larger than the $^{23}$Na cloud, the frequency of oscillation in the PRF of $^{39}$K atoms remains unchanged even in the presence of $^{23}$Na atoms. We argue that the inter-species interaction strength between the cold $^{39}$K and $^{23}$Na atoms is a function of their spatial overlap. This is because, in a 3DMOT each atom follows a trajectory due to photon scattering as well as magneto-optical force \cite{foot2005atomic, metcalf2012laser} and eventually traverse the entire cloud size. Now, only when a $^{39}$K atom is in a region of space overlapping with a cold $^{23}$Na cloud, the inter-species interaction is present. Therefore, on an average the parametrized interaction strength critically depends on the overlap of these two clouds.
 \par
 When the size of the $^{39}$K atomic cloud is much larger than the trapped $^{23}$Na atomic cloud, the dynamics of the $^{39}$K atoms remains largely unaffected by the $^{23}$Na atoms. As a result, the effect of $^{23}$Na atoms on the PRF of the $^{39}$K atoms is nearly imperceptible, as illustrated in \textbf{Fig-\ref{fig:Fully_overlapped_UD}(a)}. However, when the sizes of the $^{23}$Na and $^{39}$K atomic clouds are comparable, the presence of the $^{23}$Na atoms has a significant effect on the PRF of $^{39}$K . In this case, the oscillation frequency of the PRF of the $^{39}$K atoms increases by $\sim30$\%, reflecting a clear effect of the $^{23}$Na atoms on the dynamics of the $^{39}$K atoms. The sizes of the cold-atomic clouds are determined from the FWHM of their respective fluorescence images.

\begin{figure}[h]
    \centering
    \includegraphics[width=0.8\linewidth]{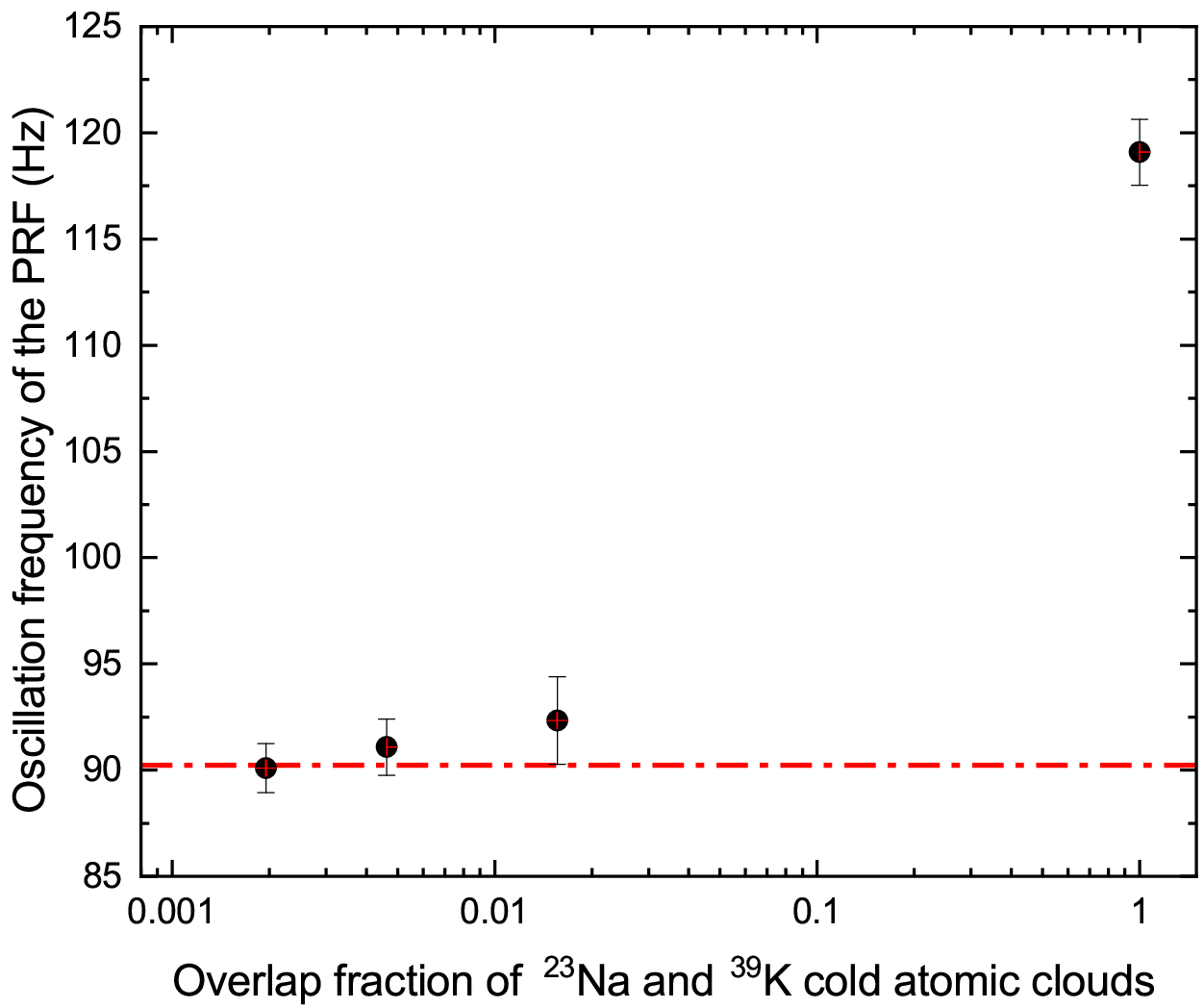}
    \caption{In the underdamped regime the oscillation frequency of the PRF has been calculated while varying the size ratio of the cold atomic clouds. The oscillation frequency goes up when the two atomic clouds are comparable in size. When only $^{39}$K atoms are present in the MOT, the oscillation frequency for the PRF of the $^{39}$K atoms is $\sim90$ Hz whereas in the presence of the $^{23}$Na atoms this frequency increases by $\sim30$\%. The dotted reference line depicts the oscillation frequency observed in the PRF when no trapped sodium atoms are present in the MOT.}
    \label{fig:Oscl_freq_vs_size_ratio}
\end{figure}

 Repeating the PRF measurements in the underdamped regime with variable overlap between the $^{39}$K and $^{23}$Na cold atomic clouds and thereby modifying the effective inter-species interaction strength, we have recorded the oscillation frequencies and presented the data in \textbf{Fig-\ref{fig:Oscl_freq_vs_size_ratio}}. 

\begin{figure}[htbp]
    \centering
    \includegraphics[width=\linewidth]{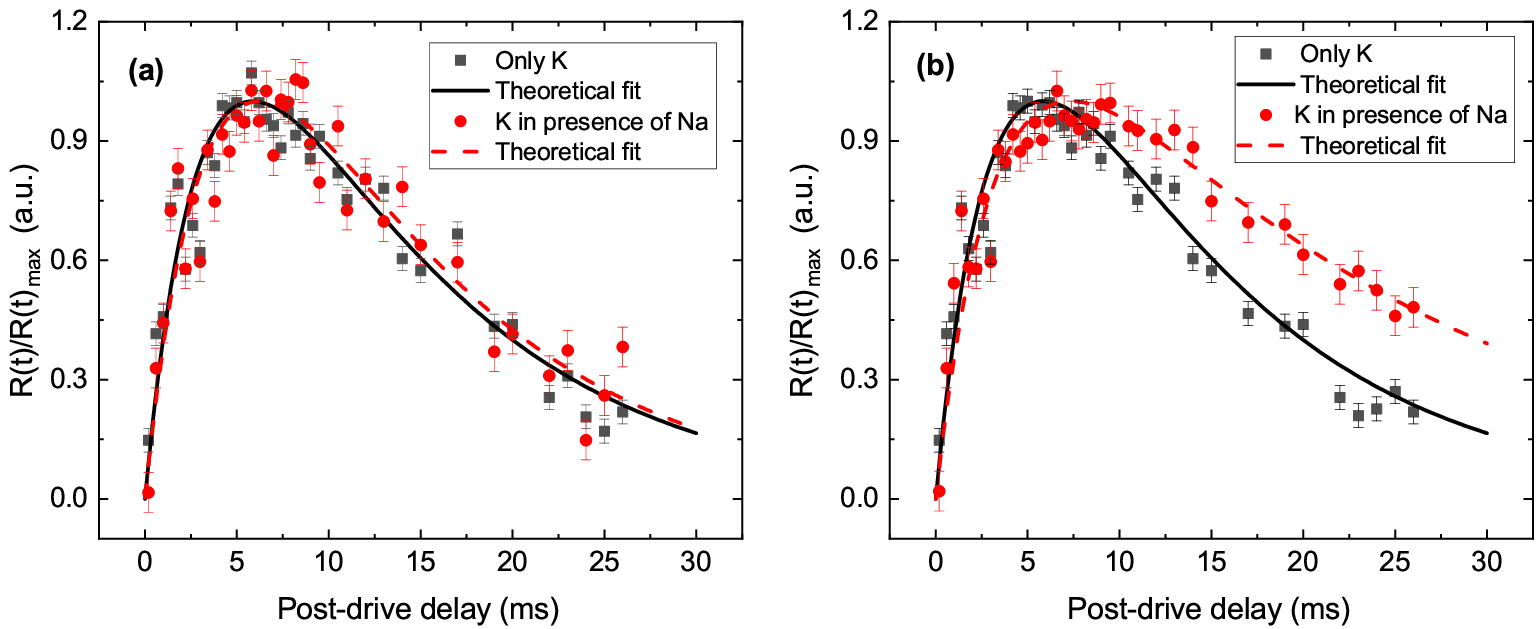}
    \caption{In the overdamped regime, the PRF of the potassium atoms is observed with and without the sodium atoms within the MOT. When the sizes of the cold atomic clouds of sodium and potassium are comparable, the damping observed in the PRF is less than the case where the size of the potassium cloud is significantly higher than that of the sodium cloud.  The parameters for the theoretical fit of normalised PRF of only $^{39}$K cloud are, $N_1 N_2 l_2=0$, $\gamma_{11}=182$ Hz, $ \gamma_{1c}=85.3$ Hz for both (a) and (b); and of $^{39}$K cloud in the presence of $^{23}$Na in case (a) $N_1 N_2 l_2=200$, $\gamma_{11}=180$ Hz, $\gamma_{22}=161$ Hz, $\gamma_{1c}=88$ Hz and $\gamma_{2c}=56$ Hz and in case (b) $N_1 N_2 l_2=980000$, $\gamma_{11}=195.5$ Hz, $\gamma_{22}=172.5$ Hz, $\gamma_{1c}=153$ Hz and $\gamma_{2c}=133$ Hz. }
    \label{fig:OD_fully_overlapped}
\end{figure}

Next, we perform another set of experiments by modifying the $^{39}$K 3DMOT parameters. In particular, the magnetic field gradient is reduced to 12 $G/cm$ and the cooling beam intensity was increased to $\sim$10.6 $I_{sat}$ (per beam) in order to make the motion of the cold atomic cloud overdamped. The entire response function measurement described above was repeated in the overdamped regime. As in the underdamped regime, in this case as well, $^{39}$K atoms have been subjected to the driving beam and then the PRF is analyzed both in the presence and absence of the trapped $^{23}$Na atoms within the MOT. While analyzing the overdamped PRF in the presence of the trapped $^{23}$Na atoms, it is observed that the damping coefficient for the PRF is decreased in this case thereby making the PRF slowly varying compared to the case where the size of the $^{39}$K atomic cloud is significantly higher than that of the $^{23}$Na atomic cloud. We observe in \textbf{Fig-\ref{fig:OD_fully_overlapped}(b)} that the damping coefficient reduces as much as 10.5$\%$ (from $\sim2.5\times10^{-23}$ kg/s to $\sim2.2\times10^{-23}$ kg/s) when the two clouds overlap completely. On the other hand, for a small fraction of overlap, the change in the
damping coefficient is not discernible. (\textbf{Fig-\ref{fig:OD_fully_overlapped}(a)})

\section{Conclusion and Outlook}
In this paper, we have experimentally and theoretically explored the position response of one species of cold atoms subjected to a sudden displacement from the equilibrium position in a magneto-optical trap (MOT), interacting with another species via light-assisted collisional channels. We find that the oscillation frequency in the trap gets significantly modified in the presence of inter-species interaction. In the overdamped regime of motion of the atoms in the cloud, the damping coefficient changes significantly as well. The interplay between these collisional effects and trap dynamics provides a pathway for exploring transport phenomena and energy dissipation in coupled atomic systems. The excellent agreement between our experimental observations and the theoretical predictions, derived from a multi-component quantum Langevin framework, highlights the robustness of this approach. These findings validate its applicability to various multi-component systems, such as electron-hole plasmas \cite{huber2001many}, ion-atom mixtures \cite{dieterle2021transport}, and hybrid quantum systems \cite{xiang2013hybrid}, where inter-component interactions fundamentally influence collective behaviors. Inter-species interactions reveal emergent phenomena, such as synchronized motions \cite{yan2024collective} and non-linear dynamics \cite{corgier2020interacting}, offering unique insights into multi-component transport properties under equilibrium and non-equilibrium conditions.  
\par
Our framework lays the groundwork for studying ultracold Bosonic mixtures near quantum phase transitions, where quantum correlations and many-body effects \cite{Alhambra_2020, Frerot_2023} become prominent. Further studies could explore 
inter-diffusion effects and miscibility and immiscibility transitions in Bose-Bose \cite{deng2024miscibility} or Bose-Fermi mixtures \cite{buchler2003supersolid}, enabling a deeper understanding of phase separation and coherent transport. Hybrid systems, such as coupled atoms and ions or Rydberg platforms \cite{dieterle2020inelastic,xing2022ion, weckesser2021observation,RMPTomza2019,Gregory_2021,Pan2024,defenu2023long,Pfau2023atom-ion,wang2020optical,hirzler2021rydberg,cheng2024emergent, dieterle2021transport,xiang2013hybrid}, present opportunities for investigating entanglement, quantum state transfer, and non-classical correlations. Additionally, the introduction of lattice potentials could uncover novel topological phenomena, including synthetic dimensions \cite{sundar2018synthetic}, topological insulators \cite{gonzalez2020dynamical}, and exotic phases of matter \cite{liu2019floquet}. These studies are expected to have implications for both fundamental physics and applications in quantum simulations. In summary, this work provides a detailed understanding of inter-species interactions in multi-component systems and their impact on transport properties, offering a platform for further exploration in quantum technology and condensed matter physics.

\section{Acknowledgment}
We would like to acknowledge the support of Ibrahim, Muneeswaran, Meena from the RRI Mechanical, Electrical and Electronics section in facilitating the completion of this experiment. Our sincere thanks to Sahana Rao and Pratik Misra for their contributions in ensuring the successful execution of the experiment.


\bibliography{revtex}

\begin{thebibliography}{37}%
\makeatletter
\providecommand \@ifxundefined [1]{%
 \@ifx{#1\undefined}
}%
\providecommand \@ifnum [1]{%
 \ifnum #1\expandafter \@firstoftwo
 \else \expandafter \@secondoftwo
 \fi
}%
\providecommand \@ifx [1]{%
 \ifx #1\expandafter \@firstoftwo
 \else \expandafter \@secondoftwo
 \fi
}%
\providecommand \natexlab [1]{#1}%
\providecommand \enquote  [1]{``#1''}%
\providecommand \bibnamefont  [1]{#1}%
\providecommand \bibfnamefont [1]{#1}%
\providecommand \citenamefont [1]{#1}%
\providecommand \href@noop [0]{\@secondoftwo}%
\providecommand \href [0]{\begingroup \@sanitize@url \@href}%
\providecommand \@href[1]{\@@startlink{#1}\@@href}%
\providecommand \@@href[1]{\endgroup#1\@@endlink}%
\providecommand \@sanitize@url [0]{\catcode `\\12\catcode `\$12\catcode `\&12\catcode `\#12\catcode `\^12\catcode `\_12\catcode `\%12\relax}%
\providecommand \@@startlink[1]{}%
\providecommand \@@endlink[0]{}%
\providecommand \url  [0]{\begingroup\@sanitize@url \@url }%
\providecommand \@url [1]{\endgroup\@href {#1}{\urlprefix }}%
\providecommand \urlprefix  [0]{URL }%
\providecommand \Eprint [0]{\href }%
\providecommand \doibase [0]{http://dx.doi.org/}%
\providecommand \selectlanguage [0]{\@gobble}%
\providecommand \bibinfo  [0]{\@secondoftwo}%
\providecommand \bibfield  [0]{\@secondoftwo}%
\providecommand \translation [1]{[#1]}%
\providecommand \BibitemOpen [0]{}%
\providecommand \bibitemStop [0]{}%
\providecommand \bibitemNoStop [0]{.\EOS\space}%
\providecommand \EOS [0]{\spacefactor3000\relax}%
\providecommand \BibitemShut  [1]{\csname bibitem#1\endcsname}%
\let\auto@bib@innerbib\@empty
\bibitem [{\citenamefont {Morawetz}(2014)}]{morawetz2014universal}%
  \BibitemOpen
  \bibfield  {author} {\bibinfo {author} {\bibfnamefont {K.}~\bibnamefont {Morawetz}},\ }\href@noop {} {\bibfield  {journal} {\bibinfo  {journal} {Physical Review B}\ }\textbf {\bibinfo {volume} {90}},\ \bibinfo {pages} {075303} (\bibinfo {year} {2014})}\BibitemShut {NoStop}%
\bibitem [{\citenamefont {Inguscio}\ and\ \citenamefont {Fallani}(2013)}]{inguscio2013atomic}%
  \BibitemOpen
  \bibfield  {author} {\bibinfo {author} {\bibfnamefont {M.}~\bibnamefont {Inguscio}}\ and\ \bibinfo {author} {\bibfnamefont {L.}~\bibnamefont {Fallani}},\ }\href@noop {} {\emph {\bibinfo {title} {Atomic Physics: Precise Measurements and Ultracold Matter}}}\ (\bibinfo  {publisher} {OUP Oxford},\ \bibinfo {year} {2013})\BibitemShut {NoStop}%
\bibitem [{\citenamefont {Bhar}\ \emph {et~al.}(2022)\citenamefont {Bhar}, \citenamefont {Swar}, \citenamefont {Satpathi}, \citenamefont {Sinha}, \citenamefont {Sorkin}, \citenamefont {Chaudhuri},\ and\ \citenamefont {Roy}}]{bhar2022measurements}%
  \BibitemOpen
  \bibfield  {author} {\bibinfo {author} {\bibfnamefont {S.}~\bibnamefont {Bhar}}, \bibinfo {author} {\bibfnamefont {M.}~\bibnamefont {Swar}}, \bibinfo {author} {\bibfnamefont {U.}~\bibnamefont {Satpathi}}, \bibinfo {author} {\bibfnamefont {S.}~\bibnamefont {Sinha}}, \bibinfo {author} {\bibfnamefont {R.}~\bibnamefont {Sorkin}}, \bibinfo {author} {\bibfnamefont {S.}~\bibnamefont {Chaudhuri}}, \ and\ \bibinfo {author} {\bibfnamefont {S.}~\bibnamefont {Roy}},\ }\href@noop {} {\bibfield  {journal} {\bibinfo  {journal} {Optics Continuum}\ }\textbf {\bibinfo {volume} {1}},\ \bibinfo {pages} {171} (\bibinfo {year} {2022})}\BibitemShut {NoStop}%
\bibitem [{\citenamefont {Misra}\ \emph {et~al.}(2024)\citenamefont {Misra}, \citenamefont {Satpathi}, \citenamefont {Sinha}, \citenamefont {Roy},\ and\ \citenamefont {Chaudhuri}}]{misra2024effect}%
  \BibitemOpen
  \bibfield  {author} {\bibinfo {author} {\bibfnamefont {A.}~\bibnamefont {Misra}}, \bibinfo {author} {\bibfnamefont {U.}~\bibnamefont {Satpathi}}, \bibinfo {author} {\bibfnamefont {S.}~\bibnamefont {Sinha}}, \bibinfo {author} {\bibfnamefont {S.}~\bibnamefont {Roy}}, \ and\ \bibinfo {author} {\bibfnamefont {S.}~\bibnamefont {Chaudhuri}},\ }\href@noop {} {\bibfield  {journal} {\bibinfo  {journal} {Optics Letters}\ }\textbf {\bibinfo {volume} {49}},\ \bibinfo {pages} {4377} (\bibinfo {year} {2024})}\BibitemShut {NoStop}%
\bibitem [{\citenamefont {Sutradhar}\ \emph {et~al.}(2023)\citenamefont {Sutradhar}, \citenamefont {Misra}, \citenamefont {Pal}, \citenamefont {Majumder}, \citenamefont {Roy},\ and\ \citenamefont {Chaudhuri}}]{sutradhar2023fast}%
  \BibitemOpen
  \bibfield  {author} {\bibinfo {author} {\bibfnamefont {S.}~\bibnamefont {Sutradhar}}, \bibinfo {author} {\bibfnamefont {A.}~\bibnamefont {Misra}}, \bibinfo {author} {\bibfnamefont {G.}~\bibnamefont {Pal}}, \bibinfo {author} {\bibfnamefont {S.}~\bibnamefont {Majumder}}, \bibinfo {author} {\bibfnamefont {S.}~\bibnamefont {Roy}}, \ and\ \bibinfo {author} {\bibfnamefont {S.}~\bibnamefont {Chaudhuri}},\ }\href@noop {} {\bibfield  {journal} {\bibinfo  {journal} {AIP Advances}\ }\textbf {\bibinfo {volume} {13}} (\bibinfo {year} {2023})}\BibitemShut {NoStop}%
\bibitem [{\citenamefont {Hewitt}\ \emph {et~al.}(2024)\citenamefont {Hewitt}, \citenamefont {Bertheas}, \citenamefont {Jain}, \citenamefont {Nishida},\ and\ \citenamefont {Barontini}}]{hewitt2024controlling}%
  \BibitemOpen
  \bibfield  {author} {\bibinfo {author} {\bibfnamefont {T.}~\bibnamefont {Hewitt}}, \bibinfo {author} {\bibfnamefont {T.}~\bibnamefont {Bertheas}}, \bibinfo {author} {\bibfnamefont {M.}~\bibnamefont {Jain}}, \bibinfo {author} {\bibfnamefont {Y.}~\bibnamefont {Nishida}}, \ and\ \bibinfo {author} {\bibfnamefont {G.}~\bibnamefont {Barontini}},\ }\href@noop {} {\bibfield  {journal} {\bibinfo  {journal} {Quantum Science and Technology}\ }\textbf {\bibinfo {volume} {9}},\ \bibinfo {pages} {035039} (\bibinfo {year} {2024})}\BibitemShut {NoStop}%
\bibitem [{\citenamefont {Bhatt}\ \emph {et~al.}(2022)\citenamefont {Bhatt}, \citenamefont {Kilinc}, \citenamefont {H{\"o}cker},\ and\ \citenamefont {Jendrzejewski}}]{bhatt2022stochastic}%
  \BibitemOpen
  \bibfield  {author} {\bibinfo {author} {\bibfnamefont {R.~P.}\ \bibnamefont {Bhatt}}, \bibinfo {author} {\bibfnamefont {J.}~\bibnamefont {Kilinc}}, \bibinfo {author} {\bibfnamefont {L.}~\bibnamefont {H{\"o}cker}}, \ and\ \bibinfo {author} {\bibfnamefont {F.}~\bibnamefont {Jendrzejewski}},\ }\href@noop {} {\bibfield  {journal} {\bibinfo  {journal} {Scientific Reports}\ }\textbf {\bibinfo {volume} {12}},\ \bibinfo {pages} {2422} (\bibinfo {year} {2022})}\BibitemShut {NoStop}%
\bibitem [{\citenamefont {Dieterle}\ \emph {et~al.}(2020)\citenamefont {Dieterle}, \citenamefont {Berngruber}, \citenamefont {H{\"o}lzl}, \citenamefont {L{\"o}w}, \citenamefont {Jachymski}, \citenamefont {Pfau},\ and\ \citenamefont {Meinert}}]{dieterle2020inelastic}%
  \BibitemOpen
  \bibfield  {author} {\bibinfo {author} {\bibfnamefont {T.}~\bibnamefont {Dieterle}}, \bibinfo {author} {\bibfnamefont {M.}~\bibnamefont {Berngruber}}, \bibinfo {author} {\bibfnamefont {C.}~\bibnamefont {H{\"o}lzl}}, \bibinfo {author} {\bibfnamefont {R.}~\bibnamefont {L{\"o}w}}, \bibinfo {author} {\bibfnamefont {K.}~\bibnamefont {Jachymski}}, \bibinfo {author} {\bibfnamefont {T.}~\bibnamefont {Pfau}}, \ and\ \bibinfo {author} {\bibfnamefont {F.}~\bibnamefont {Meinert}},\ }\href@noop {} {\bibfield  {journal} {\bibinfo  {journal} {Physical Review A}\ }\textbf {\bibinfo {volume} {102}},\ \bibinfo {pages} {041301} (\bibinfo {year} {2020})}\BibitemShut {NoStop}%
\bibitem [{\citenamefont {Xing}\ \emph {et~al.}(2022)\citenamefont {Xing}, \citenamefont {da~Silva~Jr}, \citenamefont {Vexiau}, \citenamefont {Bouloufa-Maafa}, \citenamefont {Willitsch},\ and\ \citenamefont {Dulieu}}]{xing2022ion}%
  \BibitemOpen
  \bibfield  {author} {\bibinfo {author} {\bibfnamefont {X.}~\bibnamefont {Xing}}, \bibinfo {author} {\bibfnamefont {H.}~\bibnamefont {da~Silva~Jr}}, \bibinfo {author} {\bibfnamefont {R.}~\bibnamefont {Vexiau}}, \bibinfo {author} {\bibfnamefont {N.}~\bibnamefont {Bouloufa-Maafa}}, \bibinfo {author} {\bibfnamefont {S.}~\bibnamefont {Willitsch}}, \ and\ \bibinfo {author} {\bibfnamefont {O.}~\bibnamefont {Dulieu}},\ }\href@noop {} {\bibfield  {journal} {\bibinfo  {journal} {Physical Review A}\ }\textbf {\bibinfo {volume} {106}},\ \bibinfo {pages} {062809} (\bibinfo {year} {2022})}\BibitemShut {NoStop}%
\bibitem [{\citenamefont {Weckesser}\ \emph {et~al.}(2021)\citenamefont {Weckesser}, \citenamefont {Thielemann}, \citenamefont {Wiater}, \citenamefont {Wojciechowska}, \citenamefont {Karpa}, \citenamefont {Jachymski}, \citenamefont {Tomza}, \citenamefont {Walker},\ and\ \citenamefont {Schaetz}}]{weckesser2021observation}%
  \BibitemOpen
  \bibfield  {author} {\bibinfo {author} {\bibfnamefont {P.}~\bibnamefont {Weckesser}}, \bibinfo {author} {\bibfnamefont {F.}~\bibnamefont {Thielemann}}, \bibinfo {author} {\bibfnamefont {D.}~\bibnamefont {Wiater}}, \bibinfo {author} {\bibfnamefont {A.}~\bibnamefont {Wojciechowska}}, \bibinfo {author} {\bibfnamefont {L.}~\bibnamefont {Karpa}}, \bibinfo {author} {\bibfnamefont {K.}~\bibnamefont {Jachymski}}, \bibinfo {author} {\bibfnamefont {M.}~\bibnamefont {Tomza}}, \bibinfo {author} {\bibfnamefont {T.}~\bibnamefont {Walker}}, \ and\ \bibinfo {author} {\bibfnamefont {T.}~\bibnamefont {Schaetz}},\ }\href@noop {} {\bibfield  {journal} {\bibinfo  {journal} {Nature}\ }\textbf {\bibinfo {volume} {600}},\ \bibinfo {pages} {429} (\bibinfo {year} {2021})}\BibitemShut {NoStop}%
\bibitem [{\citenamefont {Tomza}\ \emph {et~al.}(2019)\citenamefont {Tomza}, \citenamefont {Jachymski}, \citenamefont {Gerritsma}, \citenamefont {Negretti}, \citenamefont {Calarco}, \citenamefont {Idziaszek},\ and\ \citenamefont {Julienne}}]{RMPTomza2019}%
  \BibitemOpen
  \bibfield  {author} {\bibinfo {author} {\bibfnamefont {M.}~\bibnamefont {Tomza}}, \bibinfo {author} {\bibfnamefont {K.}~\bibnamefont {Jachymski}}, \bibinfo {author} {\bibfnamefont {R.}~\bibnamefont {Gerritsma}}, \bibinfo {author} {\bibfnamefont {A.}~\bibnamefont {Negretti}}, \bibinfo {author} {\bibfnamefont {T.}~\bibnamefont {Calarco}}, \bibinfo {author} {\bibfnamefont {Z.}~\bibnamefont {Idziaszek}}, \ and\ \bibinfo {author} {\bibfnamefont {P.~S.}\ \bibnamefont {Julienne}},\ }\href {\doibase 10.1103/RevModPhys.91.035001} {\bibfield  {journal} {\bibinfo  {journal} {Rev. Mod. Phys.}\ }\textbf {\bibinfo {volume} {91}},\ \bibinfo {pages} {035001} (\bibinfo {year} {2019})}\BibitemShut {NoStop}%
\bibitem [{\citenamefont {Gregory}\ \emph {et~al.}(2021)\citenamefont {Gregory}, \citenamefont {Blackmore}, \citenamefont {D}, \citenamefont {Fernley}, \citenamefont {Bromley}, \citenamefont {Hutson},\ and\ \citenamefont {Cornish}}]{Gregory_2021}%
  \BibitemOpen
  \bibfield  {author} {\bibinfo {author} {\bibfnamefont {P.~D.}\ \bibnamefont {Gregory}}, \bibinfo {author} {\bibfnamefont {J.~A.}\ \bibnamefont {Blackmore}}, \bibinfo {author} {\bibfnamefont {F.~M.}\ \bibnamefont {D}}, \bibinfo {author} {\bibfnamefont {L.~M.}\ \bibnamefont {Fernley}}, \bibinfo {author} {\bibfnamefont {S.~L.}\ \bibnamefont {Bromley}}, \bibinfo {author} {\bibfnamefont {J.~M.}\ \bibnamefont {Hutson}}, \ and\ \bibinfo {author} {\bibfnamefont {S.~L.}\ \bibnamefont {Cornish}},\ }\href {\doibase 10.1088/1367-2630/ac3c63} {\bibfield  {journal} {\bibinfo  {journal} {New Journal of Physics}\ }\textbf {\bibinfo {volume} {23}},\ \bibinfo {pages} {125004} (\bibinfo {year} {2021})}\BibitemShut {NoStop}%
\bibitem [{\citenamefont {Cao}\ \emph {et~al.}(2024)\citenamefont {Cao}, \citenamefont {Wang}, \citenamefont {Yang}, \citenamefont {Fan}, \citenamefont {Su}, \citenamefont {Rui}, \citenamefont {Zhao},\ and\ \citenamefont {Pan}}]{Pan2024}%
  \BibitemOpen
  \bibfield  {author} {\bibinfo {author} {\bibfnamefont {J.}~\bibnamefont {Cao}}, \bibinfo {author} {\bibfnamefont {B.-Y.}\ \bibnamefont {Wang}}, \bibinfo {author} {\bibfnamefont {H.}~\bibnamefont {Yang}}, \bibinfo {author} {\bibfnamefont {Z.-J.}\ \bibnamefont {Fan}}, \bibinfo {author} {\bibfnamefont {Z.}~\bibnamefont {Su}}, \bibinfo {author} {\bibfnamefont {J.}~\bibnamefont {Rui}}, \bibinfo {author} {\bibfnamefont {B.}~\bibnamefont {Zhao}}, \ and\ \bibinfo {author} {\bibfnamefont {J.-W.}\ \bibnamefont {Pan}},\ }\href {\doibase 10.1103/PhysRevLett.132.093403} {\bibfield  {journal} {\bibinfo  {journal} {Phys. Rev. Lett.}\ }\textbf {\bibinfo {volume} {132}},\ \bibinfo {pages} {093403} (\bibinfo {year} {2024})}\BibitemShut {NoStop}%
\bibitem [{\citenamefont {Defenu}\ \emph {et~al.}(2023)\citenamefont {Defenu}, \citenamefont {Donner}, \citenamefont {Macr{\`\i}}, \citenamefont {Pagano}, \citenamefont {Ruffo},\ and\ \citenamefont {Trombettoni}}]{defenu2023long}%
  \BibitemOpen
  \bibfield  {author} {\bibinfo {author} {\bibfnamefont {N.}~\bibnamefont {Defenu}}, \bibinfo {author} {\bibfnamefont {T.}~\bibnamefont {Donner}}, \bibinfo {author} {\bibfnamefont {T.}~\bibnamefont {Macr{\`\i}}}, \bibinfo {author} {\bibfnamefont {G.}~\bibnamefont {Pagano}}, \bibinfo {author} {\bibfnamefont {S.}~\bibnamefont {Ruffo}}, \ and\ \bibinfo {author} {\bibfnamefont {A.}~\bibnamefont {Trombettoni}},\ }\href@noop {} {\bibfield  {journal} {\bibinfo  {journal} {Reviews of Modern Physics}\ }\textbf {\bibinfo {volume} {95}},\ \bibinfo {pages} {035002} (\bibinfo {year} {2023})}\BibitemShut {NoStop}%
\bibitem [{\citenamefont {Zou}\ \emph {et~al.}(2023)\citenamefont {Zou}, \citenamefont {Berngruber}, \citenamefont {Anasuri}, \citenamefont {Zuber}, \citenamefont {Meinert}, \citenamefont {L\"ow},\ and\ \citenamefont {Pfau}}]{Pfau2023atom-ion}%
  \BibitemOpen
  \bibfield  {author} {\bibinfo {author} {\bibfnamefont {Y.-Q.}\ \bibnamefont {Zou}}, \bibinfo {author} {\bibfnamefont {M.}~\bibnamefont {Berngruber}}, \bibinfo {author} {\bibfnamefont {V.~S.~V.}\ \bibnamefont {Anasuri}}, \bibinfo {author} {\bibfnamefont {N.}~\bibnamefont {Zuber}}, \bibinfo {author} {\bibfnamefont {F.}~\bibnamefont {Meinert}}, \bibinfo {author} {\bibfnamefont {R.}~\bibnamefont {L\"ow}}, \ and\ \bibinfo {author} {\bibfnamefont {T.}~\bibnamefont {Pfau}},\ }\href {\doibase 10.1103/PhysRevLett.130.023002} {\bibfield  {journal} {\bibinfo  {journal} {Phys. Rev. Lett.}\ }\textbf {\bibinfo {volume} {130}},\ \bibinfo {pages} {023002} (\bibinfo {year} {2023})}\BibitemShut {NoStop}%
\bibitem [{\citenamefont {Wang}\ \emph {et~al.}(2020)\citenamefont {Wang}, \citenamefont {Dei{\ss}}, \citenamefont {Raithel},\ and\ \citenamefont {Denschlag}}]{wang2020optical}%
  \BibitemOpen
  \bibfield  {author} {\bibinfo {author} {\bibfnamefont {L.}~\bibnamefont {Wang}}, \bibinfo {author} {\bibfnamefont {M.}~\bibnamefont {Dei{\ss}}}, \bibinfo {author} {\bibfnamefont {G.}~\bibnamefont {Raithel}}, \ and\ \bibinfo {author} {\bibfnamefont {J.~H.}\ \bibnamefont {Denschlag}},\ }\href@noop {} {\bibfield  {journal} {\bibinfo  {journal} {Journal of Physics B: Atomic, Molecular and Optical Physics}\ }\textbf {\bibinfo {volume} {53}},\ \bibinfo {pages} {134005} (\bibinfo {year} {2020})}\BibitemShut {NoStop}%
\bibitem [{\citenamefont {Hirzler}\ and\ \citenamefont {P{\'e}rez-R{\'\i}os}(2021)}]{hirzler2021rydberg}%
  \BibitemOpen
  \bibfield  {author} {\bibinfo {author} {\bibfnamefont {H.}~\bibnamefont {Hirzler}}\ and\ \bibinfo {author} {\bibfnamefont {J.}~\bibnamefont {P{\'e}rez-R{\'\i}os}},\ }\href@noop {} {\bibfield  {journal} {\bibinfo  {journal} {Physical Review A}\ }\textbf {\bibinfo {volume} {103}},\ \bibinfo {pages} {043323} (\bibinfo {year} {2021})}\BibitemShut {NoStop}%
\bibitem [{\citenamefont {Cheng}\ and\ \citenamefont {Zhai}(2024)}]{cheng2024emergent}%
  \BibitemOpen
  \bibfield  {author} {\bibinfo {author} {\bibfnamefont {Y.}~\bibnamefont {Cheng}}\ and\ \bibinfo {author} {\bibfnamefont {H.}~\bibnamefont {Zhai}},\ }\href@noop {} {\bibfield  {journal} {\bibinfo  {journal} {Nature Reviews Physics}\ }\textbf {\bibinfo {volume} {6}},\ \bibinfo {pages} {566} (\bibinfo {year} {2024})}\BibitemShut {NoStop}%
\bibitem [{\citenamefont {Caldeira}\ and\ \citenamefont {Leggett}(1983)}]{Caldeira1983}%
  \BibitemOpen
  \bibfield  {author} {\bibinfo {author} {\bibfnamefont {A.}~\bibnamefont {Caldeira}}\ and\ \bibinfo {author} {\bibfnamefont {A.}~\bibnamefont {Leggett}},\ }\href {\doibase https://doi.org/10.1016/0378-4371(83)90013-4} {\bibfield  {journal} {\bibinfo  {journal} {Physica A: Statistical Mechanics and its Applications}\ }\textbf {\bibinfo {volume} {121}},\ \bibinfo {pages} {587} (\bibinfo {year} {1983})}\BibitemShut {NoStop}%
\bibitem [{\citenamefont {Ford}\ \emph {et~al.}(1988{\natexlab{a}})\citenamefont {Ford}, \citenamefont {Lewis},\ and\ \citenamefont {O'Connell}}]{PhysRevA.37.4419}%
  \BibitemOpen
  \bibfield  {author} {\bibinfo {author} {\bibfnamefont {G.~W.}\ \bibnamefont {Ford}}, \bibinfo {author} {\bibfnamefont {J.~T.}\ \bibnamefont {Lewis}}, \ and\ \bibinfo {author} {\bibfnamefont {R.~F.}\ \bibnamefont {O'Connell}},\ }\href {\doibase 10.1103/PhysRevA.37.4419} {\bibfield  {journal} {\bibinfo  {journal} {Phys. Rev. A}\ }\textbf {\bibinfo {volume} {37}},\ \bibinfo {pages} {4419} (\bibinfo {year} {1988}{\natexlab{a}})}\BibitemShut {NoStop}%
\bibitem [{\citenamefont {Ford}\ \emph {et~al.}(1988{\natexlab{b}})\citenamefont {Ford}, \citenamefont {Lewis},\ and\ \citenamefont {O'Connell}}]{ford1988independent}%
  \BibitemOpen
  \bibfield  {author} {\bibinfo {author} {\bibfnamefont {G.}~\bibnamefont {Ford}}, \bibinfo {author} {\bibfnamefont {J.}~\bibnamefont {Lewis}}, \ and\ \bibinfo {author} {\bibfnamefont {R.}~\bibnamefont {O'Connell}},\ }\href@noop {} {\bibfield  {journal} {\bibinfo  {journal} {Journal of statistical physics}\ }\textbf {\bibinfo {volume} {53}},\ \bibinfo {pages} {439} (\bibinfo {year} {1988}{\natexlab{b}})}\BibitemShut {NoStop}%
\bibitem [{\citenamefont {Jackson}(2012)}]{jackson2012classical}%
  \BibitemOpen
  \bibfield  {author} {\bibinfo {author} {\bibfnamefont {J.}~\bibnamefont {Jackson}},\ }\href {https://books.google.co.in/books?id=8qHCZjJHRUgC} {\emph {\bibinfo {title} {Classical Electrodynamics}}}\ (\bibinfo  {publisher} {Wiley},\ \bibinfo {year} {2012})\BibitemShut {NoStop}%
\bibitem [{App()}]{Appendix}%
  \BibitemOpen
  \href@noop {} {\bibfield  {journal} {\bibinfo  {journal} {ArXiv}\ }}\bibinfo {note} {See the Appendix attached to this manuscript}\BibitemShut {NoStop}%
\bibitem [{\citenamefont {Foot}(2005)}]{foot2005atomic}%
  \BibitemOpen
  \bibfield  {author} {\bibinfo {author} {\bibfnamefont {C.}~\bibnamefont {Foot}},\ }\href {https://books.google.co.in/books?id=_CoSDAAAQBAJ} {\emph {\bibinfo {title} {Atomic Physics}}},\ Oxford Master Series in Physics\ (\bibinfo  {publisher} {OUP Oxford},\ \bibinfo {year} {2005})\BibitemShut {NoStop}%
\bibitem [{\citenamefont {Metcalf}\ and\ \citenamefont {van~der Straten}(2012)}]{metcalf2012laser}%
  \BibitemOpen
  \bibfield  {author} {\bibinfo {author} {\bibfnamefont {H.}~\bibnamefont {Metcalf}}\ and\ \bibinfo {author} {\bibfnamefont {P.}~\bibnamefont {van~der Straten}},\ }\href {https://books.google.co.in/books?id=RJXwBwAAQBAJ} {\emph {\bibinfo {title} {Laser Cooling and Trapping}}},\ Graduate Texts in Contemporary Physics\ (\bibinfo  {publisher} {Springer New York},\ \bibinfo {year} {2012})\BibitemShut {NoStop}%
\bibitem [{\citenamefont {Huber}\ \emph {et~al.}(2001)\citenamefont {Huber}, \citenamefont {Tauser}, \citenamefont {Brodschelm}, \citenamefont {Bichler}, \citenamefont {Abstreiter},\ and\ \citenamefont {Leitenstorfer}}]{huber2001many}%
  \BibitemOpen
  \bibfield  {author} {\bibinfo {author} {\bibfnamefont {R.}~\bibnamefont {Huber}}, \bibinfo {author} {\bibfnamefont {F.}~\bibnamefont {Tauser}}, \bibinfo {author} {\bibfnamefont {A.}~\bibnamefont {Brodschelm}}, \bibinfo {author} {\bibfnamefont {M.}~\bibnamefont {Bichler}}, \bibinfo {author} {\bibfnamefont {G.}~\bibnamefont {Abstreiter}}, \ and\ \bibinfo {author} {\bibfnamefont {A.}~\bibnamefont {Leitenstorfer}},\ }\href@noop {} {\bibfield  {journal} {\bibinfo  {journal} {Nature}\ }\textbf {\bibinfo {volume} {414}},\ \bibinfo {pages} {286} (\bibinfo {year} {2001})}\BibitemShut {NoStop}%
\bibitem [{\citenamefont {Dieterle}\ \emph {et~al.}(2021)\citenamefont {Dieterle}, \citenamefont {Berngruber}, \citenamefont {H{\"o}lzl}, \citenamefont {L{\"o}w}, \citenamefont {Jachymski}, \citenamefont {Pfau},\ and\ \citenamefont {Meinert}}]{dieterle2021transport}%
  \BibitemOpen
  \bibfield  {author} {\bibinfo {author} {\bibfnamefont {T.}~\bibnamefont {Dieterle}}, \bibinfo {author} {\bibfnamefont {M.}~\bibnamefont {Berngruber}}, \bibinfo {author} {\bibfnamefont {C.}~\bibnamefont {H{\"o}lzl}}, \bibinfo {author} {\bibfnamefont {R.}~\bibnamefont {L{\"o}w}}, \bibinfo {author} {\bibfnamefont {K.}~\bibnamefont {Jachymski}}, \bibinfo {author} {\bibfnamefont {T.}~\bibnamefont {Pfau}}, \ and\ \bibinfo {author} {\bibfnamefont {F.}~\bibnamefont {Meinert}},\ }\href@noop {} {\bibfield  {journal} {\bibinfo  {journal} {Physical Review Letters}\ }\textbf {\bibinfo {volume} {126}},\ \bibinfo {pages} {033401} (\bibinfo {year} {2021})}\BibitemShut {NoStop}%
\bibitem [{\citenamefont {Xiang}\ \emph {et~al.}(2013)\citenamefont {Xiang}, \citenamefont {Ashhab}, \citenamefont {You},\ and\ \citenamefont {Nori}}]{xiang2013hybrid}%
  \BibitemOpen
  \bibfield  {author} {\bibinfo {author} {\bibfnamefont {Z.-L.}\ \bibnamefont {Xiang}}, \bibinfo {author} {\bibfnamefont {S.}~\bibnamefont {Ashhab}}, \bibinfo {author} {\bibfnamefont {J.}~\bibnamefont {You}}, \ and\ \bibinfo {author} {\bibfnamefont {F.}~\bibnamefont {Nori}},\ }\href@noop {} {\bibfield  {journal} {\bibinfo  {journal} {Reviews of Modern Physics}\ }\textbf {\bibinfo {volume} {85}},\ \bibinfo {pages} {623} (\bibinfo {year} {2013})}\BibitemShut {NoStop}%
\bibitem [{\citenamefont {Yan}\ \emph {et~al.}(2024)\citenamefont {Yan}, \citenamefont {Ni}, \citenamefont {Chuang}, \citenamefont {Dolgirev}, \citenamefont {Seetharam}, \citenamefont {Demler}, \citenamefont {Robens},\ and\ \citenamefont {Zwierlein}}]{yan2024collective}%
  \BibitemOpen
  \bibfield  {author} {\bibinfo {author} {\bibfnamefont {Z.~Z.}\ \bibnamefont {Yan}}, \bibinfo {author} {\bibfnamefont {Y.}~\bibnamefont {Ni}}, \bibinfo {author} {\bibfnamefont {A.}~\bibnamefont {Chuang}}, \bibinfo {author} {\bibfnamefont {P.~E.}\ \bibnamefont {Dolgirev}}, \bibinfo {author} {\bibfnamefont {K.}~\bibnamefont {Seetharam}}, \bibinfo {author} {\bibfnamefont {E.}~\bibnamefont {Demler}}, \bibinfo {author} {\bibfnamefont {C.}~\bibnamefont {Robens}}, \ and\ \bibinfo {author} {\bibfnamefont {M.}~\bibnamefont {Zwierlein}},\ }\href@noop {} {\bibfield  {journal} {\bibinfo  {journal} {Nature Physics}\ ,\ \bibinfo {pages} {1}} (\bibinfo {year} {2024})}\BibitemShut {NoStop}%
\bibitem [{\citenamefont {Corgier}\ \emph {et~al.}(2020)\citenamefont {Corgier}, \citenamefont {Loriani}, \citenamefont {Ahlers}, \citenamefont {Posso-Trujillo}, \citenamefont {Schubert}, \citenamefont {Rasel}, \citenamefont {Charron},\ and\ \citenamefont {Gaaloul}}]{corgier2020interacting}%
  \BibitemOpen
  \bibfield  {author} {\bibinfo {author} {\bibfnamefont {R.}~\bibnamefont {Corgier}}, \bibinfo {author} {\bibfnamefont {S.}~\bibnamefont {Loriani}}, \bibinfo {author} {\bibfnamefont {H.}~\bibnamefont {Ahlers}}, \bibinfo {author} {\bibfnamefont {K.}~\bibnamefont {Posso-Trujillo}}, \bibinfo {author} {\bibfnamefont {C.}~\bibnamefont {Schubert}}, \bibinfo {author} {\bibfnamefont {E.~M.}\ \bibnamefont {Rasel}}, \bibinfo {author} {\bibfnamefont {E.}~\bibnamefont {Charron}}, \ and\ \bibinfo {author} {\bibfnamefont {N.}~\bibnamefont {Gaaloul}},\ }\href@noop {} {\bibfield  {journal} {\bibinfo  {journal} {New Journal of Physics}\ }\textbf {\bibinfo {volume} {22}},\ \bibinfo {pages} {123008} (\bibinfo {year} {2020})}\BibitemShut {NoStop}%
\bibitem [{\citenamefont {Alhambra}\ \emph {et~al.}(2020)\citenamefont {Alhambra}, \citenamefont {Riddell},\ and\ \citenamefont {Garc\'{\i}a-Pintos}}]{Alhambra_2020}%
  \BibitemOpen
  \bibfield  {author} {\bibinfo {author} {\bibfnamefont {A.~M.}\ \bibnamefont {Alhambra}}, \bibinfo {author} {\bibfnamefont {J.}~\bibnamefont {Riddell}}, \ and\ \bibinfo {author} {\bibfnamefont {L.~P.}\ \bibnamefont {Garc\'{\i}a-Pintos}},\ }\href {\doibase 10.1103/PhysRevLett.124.110605} {\bibfield  {journal} {\bibinfo  {journal} {Phys. Rev. Lett.}\ }\textbf {\bibinfo {volume} {124}},\ \bibinfo {pages} {110605} (\bibinfo {year} {2020})}\BibitemShut {NoStop}%
\bibitem [{\citenamefont {Frérot}\ \emph {et~al.}(2023)\citenamefont {Frérot}, \citenamefont {Fadel},\ and\ \citenamefont {Lewenstein}}]{Frerot_2023}%
  \BibitemOpen
  \bibfield  {author} {\bibinfo {author} {\bibfnamefont {I.}~\bibnamefont {Frérot}}, \bibinfo {author} {\bibfnamefont {M.}~\bibnamefont {Fadel}}, \ and\ \bibinfo {author} {\bibfnamefont {M.}~\bibnamefont {Lewenstein}},\ }\href {\doibase 10.1088/1361-6633/acf8d7} {\bibfield  {journal} {\bibinfo  {journal} {Reports on Progress in Physics}\ }\textbf {\bibinfo {volume} {86}},\ \bibinfo {pages} {114001} (\bibinfo {year} {2023})}\BibitemShut {NoStop}%
\bibitem [{\citenamefont {Deng}\ \emph {et~al.}(2024)\citenamefont {Deng}, \citenamefont {Xue}, \citenamefont {Pang}, \citenamefont {Luo}, \citenamefont {Wang}, \citenamefont {Li},\ and\ \citenamefont {Yang}}]{deng2024miscibility}%
  \BibitemOpen
  \bibfield  {author} {\bibinfo {author} {\bibfnamefont {M.}~\bibnamefont {Deng}}, \bibinfo {author} {\bibfnamefont {M.}~\bibnamefont {Xue}}, \bibinfo {author} {\bibfnamefont {J.}~\bibnamefont {Pang}}, \bibinfo {author} {\bibfnamefont {H.}~\bibnamefont {Luo}}, \bibinfo {author} {\bibfnamefont {Z.}~\bibnamefont {Wang}}, \bibinfo {author} {\bibfnamefont {J.}~\bibnamefont {Li}}, \ and\ \bibinfo {author} {\bibfnamefont {D.}~\bibnamefont {Yang}},\ }\href@noop {} {\bibfield  {journal} {\bibinfo  {journal} {Physical Review A}\ }\textbf {\bibinfo {volume} {109}},\ \bibinfo {pages} {043324} (\bibinfo {year} {2024})}\BibitemShut {NoStop}%
\bibitem [{\citenamefont {B{\"u}chler}\ and\ \citenamefont {Blatter}(2003)}]{buchler2003supersolid}%
  \BibitemOpen
  \bibfield  {author} {\bibinfo {author} {\bibfnamefont {H.}~\bibnamefont {B{\"u}chler}}\ and\ \bibinfo {author} {\bibfnamefont {G.}~\bibnamefont {Blatter}},\ }\href@noop {} {\bibfield  {journal} {\bibinfo  {journal} {Physical review letters}\ }\textbf {\bibinfo {volume} {91}},\ \bibinfo {pages} {130404} (\bibinfo {year} {2003})}\BibitemShut {NoStop}%
\bibitem [{\citenamefont {Sundar}\ \emph {et~al.}(2018)\citenamefont {Sundar}, \citenamefont {Gadway},\ and\ \citenamefont {Hazzard}}]{sundar2018synthetic}%
  \BibitemOpen
  \bibfield  {author} {\bibinfo {author} {\bibfnamefont {B.}~\bibnamefont {Sundar}}, \bibinfo {author} {\bibfnamefont {B.}~\bibnamefont {Gadway}}, \ and\ \bibinfo {author} {\bibfnamefont {K.~R.}\ \bibnamefont {Hazzard}},\ }\href@noop {} {\bibfield  {journal} {\bibinfo  {journal} {Scientific reports}\ }\textbf {\bibinfo {volume} {8}},\ \bibinfo {pages} {1} (\bibinfo {year} {2018})}\BibitemShut {NoStop}%
\bibitem [{\citenamefont {Gonz{\'a}lez-Cuadra}\ \emph {et~al.}(2020)\citenamefont {Gonz{\'a}lez-Cuadra}, \citenamefont {Dauphin}, \citenamefont {Grzybowski}, \citenamefont {Lewenstein},\ and\ \citenamefont {Bermudez}}]{gonzalez2020dynamical}%
  \BibitemOpen
  \bibfield  {author} {\bibinfo {author} {\bibfnamefont {D.}~\bibnamefont {Gonz{\'a}lez-Cuadra}}, \bibinfo {author} {\bibfnamefont {A.}~\bibnamefont {Dauphin}}, \bibinfo {author} {\bibfnamefont {P.~R.}\ \bibnamefont {Grzybowski}}, \bibinfo {author} {\bibfnamefont {M.}~\bibnamefont {Lewenstein}}, \ and\ \bibinfo {author} {\bibfnamefont {A.}~\bibnamefont {Bermudez}},\ }\href@noop {} {\bibfield  {journal} {\bibinfo  {journal} {Physical Review Letters}\ }\textbf {\bibinfo {volume} {125}},\ \bibinfo {pages} {265301} (\bibinfo {year} {2020})}\BibitemShut {NoStop}%
\bibitem [{\citenamefont {Liu}\ \emph {et~al.}(2019)\citenamefont {Liu}, \citenamefont {Xiong}, \citenamefont {Zhang},\ and\ \citenamefont {An}}]{liu2019floquet}%
  \BibitemOpen
  \bibfield  {author} {\bibinfo {author} {\bibfnamefont {H.}~\bibnamefont {Liu}}, \bibinfo {author} {\bibfnamefont {T.-S.}\ \bibnamefont {Xiong}}, \bibinfo {author} {\bibfnamefont {W.}~\bibnamefont {Zhang}}, \ and\ \bibinfo {author} {\bibfnamefont {J.-H.}\ \bibnamefont {An}},\ }\href@noop {} {\bibfield  {journal} {\bibinfo  {journal} {Physical Review A}\ }\textbf {\bibinfo {volume} {100}},\ \bibinfo {pages} {023622} (\bibinfo {year} {2019})}\BibitemShut {NoStop}%
\end{thebibliography}%
\pagebreak

\section*{appendix}\label{appendix}
\subsection{Position response function of the first species}
The imaginary part of the PRF of the first species i.e. $^{39}$K for Ohmic bath model, is given by
\begin{align}
    \Im R_{X_c}(\omega)&=\frac{\omega\left(N_1^2 D_2 D_2^*  \gamma_1+N_1^2 N_2^2 l_2^2 \gamma_2\right)}{D D^*}\label{ImRx}
\end{align}
Here,
$$D_j=-M_j^2 \omega^2 +W_j^2-i\omega N_j\gamma_j, j=1,2, \text{and} \;D=D_1 D_2-N_1 N_2 l_2^2$$
where, $M_1\omega_1^2 -l_1 (N_1-1)=W_1^2, M_2\omega_2^2 -l_1 (N_2-1)=W_2^2$. $l_1$ and $l_2$ pertains to the strength of intra- and inter-species interaction respectively.
For finite $l_2$, the deviation from the single species PRF behaviour depends on the second term of Eq. \ref{ImRx} which is a function of $N_1N_2l_2$. In order to understand this deviation we can do a Binomial expansion regarding ${\frac{N_1N_2l_2^2}{D_1D_2}}$ and ${\frac{N_1N_2l_2^2}{D_1^*D_2^*}}$ as small parameters: 
\begin{align}
 \Im R_{X_c}(\omega)=&\frac{\omega N_1^2 D_2 D_2^* \gamma_1}{D_1D_2\left(1-\frac{N_1N_2l_2^2}{D_1D_2}\right)D_1^*D_2^*\left(1-\frac{N_1N_2l_2^2}{D_1^*D_2^*}\right)}\left(1+\frac{N_2^2 l_2^2}{D_2D_2^*}\frac{\gamma_2}{\gamma_1}\right) \nonumber\\
 =&\frac{\omega N_1^2\gamma_1}{D_1D_1^*}\left(1+\frac{N_2^2 l_2^2}{D_2D_2^*}\frac{\gamma_2}{\gamma_1}\right)
 \left(1+\frac{N_1N_2l_2^2}{D_1D_2}+\left( \frac{N_1N_2l_2^2}{D_1D_2}\right)^2+...\right)\nonumber\\&\left(1+\frac{N_1N_2l_2^2}{D_1^*D_2^*}+\left( \frac{N_1N_2l_2^2}{D_1^*D_2^*}\right)^2+...\right)\nonumber\\
 =& \frac{\omega N_1^2\gamma_1}{D_1D_1^*}+\frac{\omega N_1^2 N_2^2 l_2^2 \gamma_2}{D_1 D_1^* D_2 D_2^*}+\frac{\omega N_1^3 N_2 l_2^2 \gamma_1}{D_1^2 D_1^* D_2}+\frac{\omega N_1^3 N_2 l_2^2 \gamma_1}{D_1 D_1^{*^2} D_2}\nonumber\\
 =& \Im R_{X_c}(\omega)|_{l_2=0}+\frac{\omega N_1^2 N_2^2 l_2^2 \gamma_2}{D_1 D_1^* D_2 D_2^*}+\frac{\omega N_1^3 N_2 l_2^2 \gamma_1}{D_1^2 D_1^* D_2}+\frac{\omega N_1^3 N_2 l_2^2 \gamma_1}{D_1 D_1^{*^2} D_2}
\end{align}
The above expression is obtained truncating the expansion to the first order terms only. Using Kramers-Kronig relation we can calculate the real parts of the PRF.
Once we get the imaginary and real parts, we calculate the PRF for $^{39}$K.
Hence the time dependent PRF of $^{39}$K reduces to,

\begin{align}
   R_{X_c}(t) = & R_{X_c}(t) \big|_{l_2 = 0} + \frac{N_1^2 N_2 l_2^2}{4M_1^2 M_2} 
   \Re\Bigg[\frac{e^{-\gamma_{11}t}}{Dc_1}\Big(\gamma_{11}\gamma_{1c} C_1 \cosh(\gamma_{1c}t) + C_2 \sinh(\gamma_{1c}t)\Big) \nonumber \\
   & + \frac{4 e^{-\gamma_{22}t}}{Dc_2}\Big(\gamma_{2c} C_3 \cosh(\gamma_{2c}t) - \gamma_{22} C_4 \sinh(\gamma_{2c}t)\Big)\Bigg] 
   \label{Rtfinal}
\end{align}
where, $\gamma_{11}=\frac{N_1\gamma_1}{2 M_1}$, $\gamma_{22}=\frac{N_2\gamma_1}{2 M_2}$, $\gamma_{1c}=\sqrt{\gamma_{11}^2-W_{11}^2}$, $\gamma_{2c}=\sqrt{\gamma_{22}^2-W_{22}^2}$ and $W_{11}=W_1/\sqrt{M_1}$, $W_{22}=W_2/\sqrt{M_2}$. $\gamma_{11},\gamma_{22}$ are the damping parameters and $W_{11}, W_{22}$ are the oscillation frequencies of $^{39}$K (first species) and $^{23}$Na, respectively.
Here $C1, C2, Dc_1, C3, C4, Dc_2$ are given by, 
\begin{align}
    C1&=-\gamma_{11}^8 t \left(3 \gamma_{1c}^2+3 \gamma_{22}^2+5 \gamma_{2c}^2\right)+8 \gamma_{11}^7 \left(\gamma_{1c}^2 (\gamma_{22} t+1)+\gamma_{22} t \left(\gamma_{22}^2+\gamma_{2c}^2\right)\right)\nonumber\\
    &+2 \gamma_{11}^6 \left(\gamma_{1c}^2 \left(-2 \gamma_{22}^2 t+8 \gamma_{22}+2 \gamma_{2c}^2 t\right)+\gamma_{1c}^4 t+t \left(2 \gamma_{22}^2 \gamma_{2c}^2+\gamma_{22}^4+5 \gamma_{2c}^4\right)\right)\nonumber\\
    &-4 \gamma_{11}^5 \left(2 \gamma_{1c}^2 \left(\gamma_{22}^3 t+7 \gamma_{22}^2+\gamma_{22} \gamma_{2c}^2 t+\gamma_{2c}^2\right)+3 \gamma_{1c}^4 (\gamma_{22} t+2)\right.\nonumber\\
    &\left.+\gamma_{22} t \left(2 \gamma_{22}^2 \gamma_{2c}^2+3 \gamma_{22}^4+3 \gamma_{2c}^4\right)\right)+2 \gamma_{11}^4 \left(\gamma_{1c}^4 \left(7 \gamma_{22}^2 t-8 \gamma_{22}+\gamma_{2c}^2 t\right)\right.\nonumber\\
    &+\gamma_{1c}^2 \left(-18 \gamma_{22}^2 \gamma_{2c}^2 t+7 \gamma_{22}^4 t+8 \gamma_{22}^3+8 \gamma_{22} \gamma_{2c}^2+3 \gamma_{2c}^4 t\right)+\gamma_{1c}^6 t\nonumber\\
    &\left.+t \left(\gamma_{22}^4 \gamma_{2c}^2+3 \gamma_{22}^2 \gamma_{2c}^4+\gamma_{22}^6-5 \gamma_{2c}^6\right)\right)+8 \gamma_{11}^3 \left(-\gamma_{1c}^4 \left(\gamma_{22}^3 t-2 \gamma_{22}^2+\gamma_{22} \gamma_{2c}^2 t+2 \gamma_{2c}^2\right)\right.\nonumber\\
    &-\gamma_{1c}^2 \left(-10 \gamma_{22}^3 \gamma_{2c}^2 t-14 \gamma_{22}^2 \gamma_{2c}^2+\gamma_{22}^5 t-3 \gamma_{22}^4+\gamma_{22} \gamma_{2c}^4 t+\gamma_{2c}^4\right)+\gamma_{1c}^6 (\gamma_{22} t+3)\nonumber\\
    &\left.+\gamma_{22} t \left(\gamma_{22}^2-\gamma_{2c}^2\right)^2 \left(\gamma_{22}^2+\gamma_{2c}^2\right)\right)-\gamma_{11}^2 \left(4 \gamma_{1c}^6 \left(\gamma_{22}^2 t+4 \gamma_{22}-\gamma_{2c}^2 t\right)\right.\nonumber\\
    &-2 \gamma_{1c}^4 \left(-18 \gamma_{22}^2 \gamma_{2c}^2 t+7 \gamma_{22}^4 t+16 \gamma_{22}^3+16 \gamma_{22} \gamma_{2c}^2+3 \gamma_{2c}^4 t\right)\nonumber\\
    &+4 \gamma_{1c}^2 \left(9 \gamma_{22}^4 \gamma_{2c}^2 t-13 \gamma_{22}^2 \gamma_{2c}^4 t+24 \gamma_{22}^3 \gamma_{2c}^2+\gamma_{22}^6 t+4 \gamma_{22}^5+4 \gamma_{22} \gamma_{2c}^4+3 \gamma_{2c}^6 t\right)+3 \gamma_{1c}^8 t\nonumber\\
    &\left.+t \left(\gamma_{22}^2-\gamma_{2c}^2\right)^3 \left(3 \gamma_{22}^2+5 \gamma_{2c}^2\right)\right)-2 \gamma_{11}^9 \gamma_{22} t+\gamma_{11}^{10} t-2 \gamma_{11} \left(-2 \gamma_{1c}^2 \left(\gamma_{22}^2+\gamma_{2c}^2\right)+\gamma_{1c}^4+\left(\gamma_{22}^2-\gamma_{2c}^2\right)^2\right)\nonumber\\
    &\left(-2 \gamma_{1c}^2 \left(\gamma_{22}^3 t+6 \gamma_{22}^2+\gamma_{22} \gamma_{2c}^2 t+2 \gamma_{2c}^2\right)+\gamma_{1c}^4 (\gamma_{22} t+4)+\gamma_{22} t \left(\gamma_{22}^2-\gamma_{2c}^2\right)^2\right)+\nonumber\\
    &\left(-2 \gamma_{1c}^2 \left(\gamma_{22}^2+\gamma_{2c}^2\right)+\gamma_{1c}^4+\left(\gamma_{22}^2-\gamma_{2c}^2\right)^2\right) \left(-\gamma_{1c}^4 \left(\gamma_{22}^2 t-16 \gamma_{22}+3 \gamma_{2c}^2 t\right)-\gamma_{1c}^2 \right.\nonumber\\
    &\left.\left(2 \gamma_{22}^2 \gamma_{2c}^2 t+\gamma_{22}^4 t+16 \gamma_{22}^3+16 \gamma_{22} \gamma_{2c}^2-3 \gamma_{2c}^4 t\right)+\gamma_{1c}^6 t+t \left(\gamma_{22}^2-\gamma_{2c}^2\right)^3\right)
\end{align}
    \begin{align}
   C2&=-\gamma_{11}^{11}+2 \left(t \gamma_{1c}^2+\gamma_{22}\right) \gamma_{11}^{10}+\left((9-2 \gamma_{22} t) \gamma_{1c}^2+3 \gamma_{22}^2+5 \gamma_{2c}^2\right) \gamma_{11}^9-2 \left(4 t \gamma_{1c}^4\right.\nonumber\\
   &\left.+\left(4 t \gamma_{22}^2+3 \gamma_{22}+4 \gamma_{2c}^2 t\right) \gamma_{1c}^2+4 \gamma_{22} \left(\gamma_{22}^2+\gamma_{2c}^2\right)\right) \gamma_{11}^8+2 \left((4 \gamma_{22} t-9) \gamma_{1c}^4+\right.\nonumber\\
   &\left.2 \left(2 t \gamma_{22}^3-7 \gamma_{22}^2+2 \gamma_{2c}^2 t \gamma_{22}-5 \gamma_{2c}^2\right) \gamma_{1c}^2-\gamma_{22}^4-5 \gamma_{2c}^4-2 \gamma_{22}^2 \gamma_{2c}^2\right) \gamma_{11}^7+4 \left(3 t \gamma_{1c}^6+\right.\nonumber\\
   &\left(2 t \gamma_{22}^2+9 \gamma_{22}+2 \gamma_{2c}^2 t\right) \gamma_{1c}^4+\left(3 t \gamma_{22}^4+4 \gamma_{22}^3+2 \gamma_{2c}^2 t \gamma_{22}^2+4 \gamma_{2c}^2 \gamma_{22}+3 \gamma_{2c}^4 t\right) \gamma_{1c}^2+\nonumber\\
   &\left.3 \gamma_{22}^5+3 \gamma_{22} \gamma_{2c}^4+2 \gamma_{22}^3 \gamma_{2c}^2\right) \gamma_{11}^6-2 \left((6 \gamma_{22} t-5) \gamma_{1c}^6+\left(4 t \gamma_{22}^3+19 \gamma_{22}^2+4 \gamma_{2c}^2 t \gamma_{22}-3 \gamma_{2c}^2\right) \gamma_{1c}^4\right.\nonumber\\
   &\left.+\left(6 t \gamma_{22}^5-15 \gamma_{22}^4+4 \gamma_{2c}^2 t \gamma_{22}^3-38 \gamma_{2c}^2 \gamma_{22}^2+6 \gamma_{2c}^4 t \gamma_{22}-3 \gamma_{2c}^4\right) \gamma_{1c}^2+\gamma_{22}^6-5 \gamma_{2c}^6+3 \gamma_{22}^2 \gamma_{2c}^4+\gamma_{22}^4 \gamma_{2c}^2\right) \gamma_{11}^5\nonumber\\
   &-4 \left(2 t \gamma_{1c}^8+\left(-2 t \gamma_{22}^2+15 \gamma_{22}-2 \gamma_{2c}^2 t\right) \gamma_{1c}^6-2 \left(t \gamma_{22}^4+6 \gamma_{22}^3-10 \gamma_{2c}^2 t \gamma_{22}^2+6 \gamma_{2c}^2 \gamma_{22}+\gamma_{2c}^4 t\right) \gamma_{1c}^4\right.\nonumber\\
   &+\left(2 t \gamma_{22}^6+3 \gamma_{22}^5-2 \gamma_{2c}^2 t \gamma_{22}^4+34 \gamma_{2c}^2 \gamma_{22}^3-2 \gamma_{2c}^4 t \gamma_{22}^2+3 \gamma_{2c}^4 \gamma_{22}+2 \gamma_{2c}^6 t\right) \gamma_{1c}^2+2 \gamma_{22} \left(\gamma_{22}^2-\gamma_{2c}^2\right)^2 \nonumber\\   &\left.\left(\gamma_{22}^2+\gamma_{2c}^2\right)\right) \gamma_{11}^4+\left((8 \gamma_{22} t+3) \gamma_{1c}^8-4 \left(2 t \gamma_{22}^3-13 \gamma_{22}^2+2 \gamma_{2c}^2 t \gamma_{22}+\gamma_{2c}^2\right) \gamma_{1c}^6-\right.\nonumber\\
   &2 \left(4 t \gamma_{22}^5+23 \gamma_{22}^4-40 \gamma_{2c}^2 t \gamma_{22}^3-2 \gamma_{2c}^2 \gamma_{22}^2+4 \gamma_{2c}^4 t \gamma_{22}+3 \gamma_{2c}^4\right) \gamma_{1c}^4+4 \left(\gamma_{22}^2-\gamma_{2c}^2\right)\nonumber\\
   &\left.\left(2 t \gamma_{22}^5-3 \gamma_{22}^4+14 \gamma_{2c}^2 \gamma_{22}^2-2 \gamma_{2c}^4 t \gamma_{22}-3 \gamma_{2c}^4\right) \gamma_{1c}^2+\left(\gamma_{22}^2-\gamma_{2c}^2\right)^3 \left(3 \gamma_{22}^2+5 \gamma_{2c}^2\right)\right) \gamma_{11}^3\nonumber\\
   &+2 \left(t \gamma_{1c}^{10}+\left(-4 t \gamma_{22}^2+13 \gamma_{22}-4 \gamma_{2c}^2 t\right) \gamma_{1c}^8+\left(6 t \gamma_{22}^4-24 \gamma_{22}^3+4 \gamma_{2c}^2 t \gamma_{22}^2-24 \gamma_{2c}^2 \gamma_{22}+6 \gamma_{2c}^4 t\right) \gamma_{1c}^6\right.\nonumber\\
   &+\left(-4 t \gamma_{22}^6+10 \gamma_{22}^5+4 \gamma_{2c}^2 t \gamma_{22}^4+28 \gamma_{2c}^2 \gamma_{22}^3+4 \gamma_{2c}^4 t \gamma_{22}^2+10 \gamma_{2c}^4 \gamma_{22}-4 \gamma_{2c}^6 t\right) \gamma_{1c}^4+\left(\gamma_{22}^2-\gamma_{2c}^2\right)^4 t \gamma_{1c}^2\nonumber\\
   &\left.+\gamma_{22} \left(\gamma_{22}^2-\gamma_{2c}^2\right)^4\right) \gamma_{11}^2-\left(\gamma_{1c}^4-2 \left(\gamma_{22}^2+\gamma_{2c}^2\right) \gamma_{1c}^2+\left(\gamma_{22}^2-\gamma_{2c}^2\right)^2\right)^2 \left((2 \gamma_{22} t+3) \gamma_{1c}^2+\gamma_{22}^2-\gamma_{2c}^2\right) \gamma_{11}\nonumber\\
   &+2 \gamma_{1c}^2 \gamma_{22} \left(\gamma_{1c}^4-2 \left(\gamma_{22}^2+\gamma_{2c}^2\right) \gamma_{1c}^2+\left(\gamma_{22}^2-\gamma_{2c}^2\right)^2\right)^2
   \end{align}
   \begin{align}
   Dc_1&=\gamma_{11} \gamma_{1c}^3 (\gamma_{11}+\gamma_{1c}-\gamma_{22}-\gamma_{2c})^2 (\gamma_{11}-\gamma_{1c}+\gamma_{22}-\gamma_{2c}) (\gamma_{11}+\gamma_{1c}+\gamma_{22}-\gamma_{2c}) (\gamma_{11}-\gamma_{1c}-\gamma_{22}+\gamma_{2c})^2 \nonumber\\
   &(\gamma_{11}+\gamma_{1c}-\gamma_{22}+\gamma_{2c})^2 (\gamma_{11}-\gamma_{1c}+\gamma_{22}+\gamma_{2c}) (-\gamma_{11}+\gamma_{1c}+\gamma_{22}+\gamma_{2c})^2 (\gamma_{11}+\gamma_{1c}+\gamma_{22}+\gamma_{2c}); 
   \end{align}
   \begin{align}
    C3&=-2 \gamma_{11}^9 \left(5 \gamma_{1c}^2-7 \gamma_{22}^2+3 \gamma_{2c}^2\right)-\gamma_{11}^8 \gamma_{22} \left(3 \gamma_{1c}^2+5 \left(\gamma_{22}^2+\gamma_{2c}^2\right)\right)+4 \gamma_{11}^7 \left(2 \gamma_{1c}^2 \left(\gamma_{2c}^2-5 \gamma_{22}^2\right)+5 \gamma_{1c}^4\right.\nonumber\\
    &\left.+10 \gamma_{22}^2 \gamma_{2c}^2-11 \gamma_{22}^4+\gamma_{2c}^4\right)+2 \gamma_{11}^6 \gamma_{22} \left(2 \gamma_{1c}^2 \left(\gamma_{22}^2+\gamma_{2c}^2\right)+\gamma_{1c}^4+6 \gamma_{22}^2 \gamma_{2c}^2+5 \gamma_{22}^4+5 \gamma_{2c}^4\right)-4 \gamma_{11}^5 \nonumber\\
    &\left(-3 \gamma_{1c}^4 \left(3 \gamma_{22}^2+\gamma_{2c}^2\right)-\gamma_{1c}^2 \left(-6 \gamma_{22}^2 \gamma_{2c}^2+5 \gamma_{22}^4+\gamma_{2c}^4\right)+5 \gamma_{1c}^6+27 \gamma_{22}^2 \gamma_{2c}^4-19 \gamma_{22}^4 \gamma_{2c}^2-7 \gamma_{22}^6-\gamma_{2c}^6\right)\nonumber\\
    &+2 \gamma_{11}^4 \gamma_{22} \left(\gamma_{1c}^4 \left(\gamma_{22}^2+\gamma_{2c}^2\right)+\gamma_{1c}^2 \left(-22 \gamma_{22}^2 \gamma_{2c}^2+3 \gamma_{22}^4+3 \gamma_{2c}^4\right)+\gamma_{1c}^6-3 \gamma_{22}^2 \gamma_{2c}^4-3 \gamma_{22}^4 \gamma_{2c}^2-5 \gamma_{22}^6-5 \gamma_{2c}^6\right)\nonumber\\
    &+2 \gamma_{11}^3 \left(-4 \gamma_{1c}^6 \left(\gamma_{22}^2+3 \gamma_{2c}^2\right)-2 \gamma_{1c}^4 \left(2 \gamma_{22}^2 \gamma_{2c}^2+\gamma_{22}^4-3 \gamma_{2c}^4\right)-4 \gamma_{1c}^2 \left(\gamma_{22}^2-\gamma_{2c}^2\right)^3+5 \gamma_{1c}^8+20 \gamma_{22}^2 \gamma_{2c}^6\right.\nonumber\\
    &\left.+38 \gamma_{22}^4 \gamma_{2c}^4-60 \gamma_{22}^6 \gamma_{2c}^2+5 \gamma_{22}^8-3 \gamma_{2c}^8\right)+\gamma_{11}^2 \gamma_{22} \left(4 \gamma_{1c}^6 \left(\gamma_{22}^2+\gamma_{2c}^2\right)+\gamma_{1c}^4 \left(-44 \gamma_{22}^2 \gamma_{2c}^2+6 \gamma_{22}^4+6 \gamma_{2c}^4\right)\right.\nonumber\\
    &\left.-4 \gamma_{1c}^2 \left(-11 \gamma_{22}^4 \gamma_{2c}^2-11 \gamma_{22}^2 \gamma_{2c}^4+3 \gamma_{22}^6+3 \gamma_{2c}^6\right)-3 \gamma_{1c}^8+\left(\gamma_{22}^2-\gamma_{2c}^2\right)^2 \left(6 \gamma_{22}^2 \gamma_{2c}^2+5 \gamma_{22}^4+5 \gamma_{2c}^4\right)\right)\nonumber\\
    &+\gamma_{11}^{10} \gamma_{22}+2 \gamma_{11}^{11}-2 \gamma_{11} \left(\gamma_{1c}^8 \left(\gamma_{22}^2-5 \gamma_{2c}^2\right)+\gamma_{1c}^6 \left(4 \gamma_{22}^2 \gamma_{2c}^2-14 \gamma_{22}^4+10 \gamma_{2c}^4\right)+2 \gamma_{1c}^4 \left(\gamma_{22}^4 \gamma_{2c}^2-9 \gamma_{22}^2 \gamma_{2c}^4\right.\right.\nonumber\\
    &\left.+13 \gamma_{22}^6-5 \gamma_{2c}^6\right)+\gamma_{1c}^2 \left(4 \gamma_{22}^6 \gamma_{2c}^2-10 \gamma_{22}^4 \gamma_{2c}^4+20 \gamma_{22}^2 \gamma_{2c}^6-19 \gamma_{22}^8+5 \gamma_{2c}^8\right)+\gamma_{1c}^{10}+\left(\gamma_{22}^2-\gamma_{2c}^2\right)^3 \nonumber\\
    &\left.\left(10 \gamma_{22}^2 \gamma_{2c}^2+5 \gamma_{22}^4+\gamma_{2c}^4\right)\right)-\gamma_{22} \left(-\gamma_{1c}^2+\gamma_{22}^2+\gamma_{2c}^2\right) \left(-2 \gamma_{1c}^2 \left(\gamma_{22}^2+\gamma_{2c}^2\right)+\gamma_{1c}^4+\left(\gamma_{22}^2-\gamma_{2c}^2\right)^2\right)^2;
    \end{align}
    \begin{align}
    C4&=-2 \gamma_{11}^9 \left(5 \gamma_{1c}^2+3 \gamma_{22}^2-7 \gamma_{2c}^2\right)-\gamma_{11}^8 \gamma_{2c} \left(3 \gamma_{1c}^2+5 \left(\gamma_{22}^2+\gamma_{2c}^2\right)\right)+4 \gamma_{11}^7 \left(2 \gamma_{1c}^2 \left(\gamma_{22}^2-5 \gamma_{2c}^2\right)+5 \gamma_{1c}^4\right.\nonumber\\
    &\left.+10 \gamma_{22}^2 \gamma_{2c}^2+\gamma_{22}^4-11 \gamma_{2c}^4\right)+2 \gamma_{11}^6 \gamma_{2c} \left(2 \gamma_{1c}^2 \left(\gamma_{22}^2+\gamma_{2c}^2\right)+\gamma_{1c}^4+6 \gamma_{22}^2 \gamma_{2c}^2+5 \gamma_{22}^4+5 \gamma_{2c}^4\right)-4 \gamma_{11}^5 \nonumber\\
    &\left(-3 \gamma_{1c}^4 \left(\gamma_{22}^2+3 \gamma_{2c}^2\right)-\gamma_{1c}^2 \left(-6 \gamma_{22}^2 \gamma_{2c}^2+\gamma_{22}^4+5 \gamma_{2c}^4\right)+5 \gamma_{1c}^6-19 \gamma_{22}^2 \gamma_{2c}^4+27 \gamma_{22}^4 \gamma_{2c}^2-\gamma_{22}^6-7 \gamma_{2c}^6\right)\nonumber\\
    &+2 \gamma_{11}^4 \gamma_{2c} \left(\gamma_{1c}^4 \left(\gamma_{22}^2+\gamma_{2c}^2\right)+\gamma_{1c}^2 \left(-22 \gamma_{22}^2 \gamma_{2c}^2+3 \gamma_{22}^4+3 \gamma_{2c}^4\right)+\gamma_{1c}^6-3 \gamma_{22}^2 \gamma_{2c}^4-3 \gamma_{22}^4 \gamma_{2c}^2-5 \gamma_{22}^6-5 \gamma_{2c}^6\right)\nonumber\\
    &+2 \gamma_{11}^3 \left(-4 \gamma_{1c}^6 \left(3 \gamma_{22}^2+\gamma_{2c}^2\right)+\gamma_{1c}^4 \left(-4 \gamma_{22}^2 \gamma_{2c}^2+6 \gamma_{22}^4-2 \gamma_{2c}^4\right)+4 \gamma_{1c}^2 \left(\gamma_{22}^2-\gamma_{2c}^2\right)^3+5 \gamma_{1c}^8-60 \gamma_{22}^2 \gamma_{2c}^6\right.\nonumber\\
    &\left.+ 38 \gamma_{22}^4 \gamma_{2c}^4+20 \gamma_{22}^6 \gamma_{2c}^2-3 \gamma_{22}^8+5 \gamma_{2c}^8\right)+\gamma_{11}^2 \gamma_{2c} \left(4 \gamma_{1c}^6 \left(\gamma_{22}^2+\gamma_{2c}^2\right)+\gamma_{1c}^4 \left(-44 \gamma_{22}^2 \gamma_{2c}^2+6 \gamma_{22}^4+6 \gamma_{2c}^4\right)\right.\nonumber\\
    &\left.-4 \gamma_{1c}^2 \left(-11 \gamma_{22}^4 \gamma_{2c}^2-11 \gamma_{22}^2 \gamma_{2c}^4+3 \gamma_{22}^6+3 \gamma_{2c}^6\right)-3 \gamma_{1c}^8+\left(\gamma_{22}^2-\gamma_{2c}^2\right)^2 \left(6 \gamma_{22}^2 \gamma_{2c}^2+5 \gamma_{22}^4+5 \gamma_{2c}^4\right)\right)\nonumber\\
    &+\gamma_{11}^{10} \gamma_{2c}+2 \gamma_{11}^{11}-2 \gamma_{11} \left(\gamma_{1c}^8 \left(\gamma_{2c}^2-5 \gamma_{22}^2\right)+2 \gamma_{1c}^6 \left(2 \gamma_{22}^2 \gamma_{2c}^2+5 \gamma_{22}^4-7 \gamma_{2c}^4\right)-2 \gamma_{1c}^4 \left(9 \gamma_{22}^4 \gamma_{2c}^2-\gamma_{22}^2 \gamma_{2c}^4\right.\right.\nonumber\\
    &+\left.5 \gamma_{22}^6-13 \gamma_{2c}^6\right)+\gamma_{1c}^2 \left(20 \gamma_{22}^6 \gamma_{2c}^2-10 \gamma_{22}^4 \gamma_{2c}^4+4 \gamma_{22}^2 \gamma_{2c}^6+5 \gamma_{22}^8-19 \gamma_{2c}^8\right)+\gamma_{1c}^{10}-\left(\gamma_{22}^2-\gamma_{2c}^2\right)^3 \nonumber\\
    &\left.\left(10 \gamma_{22}^2 \gamma_{2c}^2+\gamma_{22}^4+5 \gamma_{2c}^4\right)\right)-\gamma_{2c} \left(-\gamma_{1c}^2+\gamma_{22}^2+\gamma_{2c}^2\right) \left(-2 \gamma_{1c}^2 \left(\gamma_{22}^2+\gamma_{2c}^2\right)+\gamma_{1c}^4+\left(\gamma_{22}^2-\gamma_{2c}^2\right)^2\right)^2;
    \end{align}
    \begin{align}
    Dc_2&=\gamma_{2c} (\gamma_{11}+\gamma_{1c}-\gamma_{22}-\gamma_{2c})^2 (\gamma_{11}-\gamma_{1c}+\gamma_{22}-\gamma_{2c})^2 (\gamma_{11}+\gamma_{1c}+\gamma_{22}-\gamma_{2c})^2 (\gamma_{11}-\gamma_{1c}-\gamma_{22}+\gamma_{2c})^2\nonumber\\
    &(\gamma_{11}+\gamma_{1c}-\gamma_{22}+\gamma_{2c})^2 (\gamma_{11}-\gamma_{1c}+\gamma_{22}+\gamma_{2c})^2 (-\gamma_{11}+\gamma_{1c}+\gamma_{22}+\gamma_{2c})^2 (\gamma_{11}+\gamma_{1c}+\gamma_{22}+\gamma_{2c})^2;
\end{align}

\end{document}